\documentclass[aps,prx,reprint,superscriptaddress,nofootinbib]{revtex4-1}
\pdfoutput=1
\usepackage{color}
\usepackage[normalem]{ulem}
\usepackage{graphicx}
\usepackage{amsmath}
\usepackage{amsfonts}
\usepackage{amssymb}
\usepackage{microtype}
\usepackage{verbatim}
\usepackage{bbold}
\usepackage{hyperref}
\allowdisplaybreaks
\usepackage{multirow}
\usepackage{bbm}
\usepackage{bm}
\usepackage{braket}
\usepackage{soul}
\usepackage{placeins}

\hypersetup{
  colorlinks   = true, 
  urlcolor     = blue, 
  linkcolor    = blue, 
  citecolor   = blue 
}

\usepackage{textcomp}

\usepackage[smallerops]{newtxmath}

\newcommand{\exact}{product }

\DeclareMathOperator{\Tr}{Tr}

\begin{document}

\author{Tamiro Villazon}
\email{rtvs@bu.edu}
\affiliation{Department of Physics, Boston University, 590 Commonwealth Ave., Boston, MA 02215, USA}

\author{Pieter W. Claeys}
\affiliation{TCM Group, Cavendish Laboratory, University of Cambridge, Cambridge CB3 0HE, UK}

\author{Mohit Pandey}
\affiliation{Department of Physics, Boston University, 590 Commonwealth Ave., Boston, MA 02215, USA}

\author{Anatoli Polkovnikov}
\affiliation{Department of Physics, Boston University, 590 Commonwealth Ave., Boston, MA 02215, USA}

\author{Anushya Chandran}
\affiliation{Department of Physics, Boston University, 590 Commonwealth Ave., Boston, MA 02215, USA}

\title{Persistent dark  states in anisotropic central spin models}

\begin{abstract}

Long-lived dark states, in which an experimentally accessible qubit is not in thermal equilibrium with a surrounding spin bath, are pervasive in solid-state systems.
We explain the ubiquity of dark states in a large class of inhomogenous central spin models using the proximity to integrable lines with exact dark eigenstates.
At numerically accessible sizes, dark states persist as eigenstates at large deviations from integrability, and the qubit retains memory of its initial polarization at long times.
Although the eigenstates of the system are chaotic, exhibiting exponential sensitivity to small perturbations, they do not satisfy the eigenstate thermalization hypothesis. Rather, we predict long relaxation times that increase exponentially with system size. 
We propose that this intermediate \emph{chaotic but non-ergodic} regime characterizes mesoscopic quantum dot and diamond defect systems, as we see no numerical tendency towards conventional thermalization with a finite relaxation time.
\end{abstract}

\maketitle

\section{Introduction}
\label{sec:introduction}
State-of-the-art quantum technologies can control and coherently manipulate qubit systems with exquisite precision~\cite{awschalom2013quantum,wendin2017quantum,wineland2009quantum,vandersypen2005nmr}. 
The surrounding environment of the qubit however eventually decoheres the qubit and limits quantum applications~\cite{koch2016controlling,zhang2007modelling,lidar2014review}. An efficient way of extending coherence times is to prepare the system in so-called dark states, in which the qubit is effectively decoupled from the bath~\cite{taylor_controlling_2003,kurucz2009qubit,villazon2020integrability}. 
Dark states have been identified in several integrable central spin models~\cite{taylor_controlling_2003,villazon2020integrability}, and are central to quantum computing~\cite{niknam2018dynamics,tran_blind_2018}, metrology~\cite{sushkov_magnetic_2014,laraoui2011diamond} and control~\cite{dobrovitski2013quantum,ramsay2010review} applications in a variety of experimental qubit systems, including nitrogen vacancy centers in diamond~\cite{hall2014analytic,rios2010quantum} and semiconducting quantum dots~\cite{hanson_review_2007,schliemann2003electron}. 
A central goal of this work is to show that dark states can persist in experimentally relevant non-integrable central spin models. At numerically accessible system sizes, they exist as exact eigenstates. In the thermodynamic limit, the qubit could eventually thermalize but only after long times.

\begin{figure}[tp]
\includegraphics[width=0.8\columnwidth]{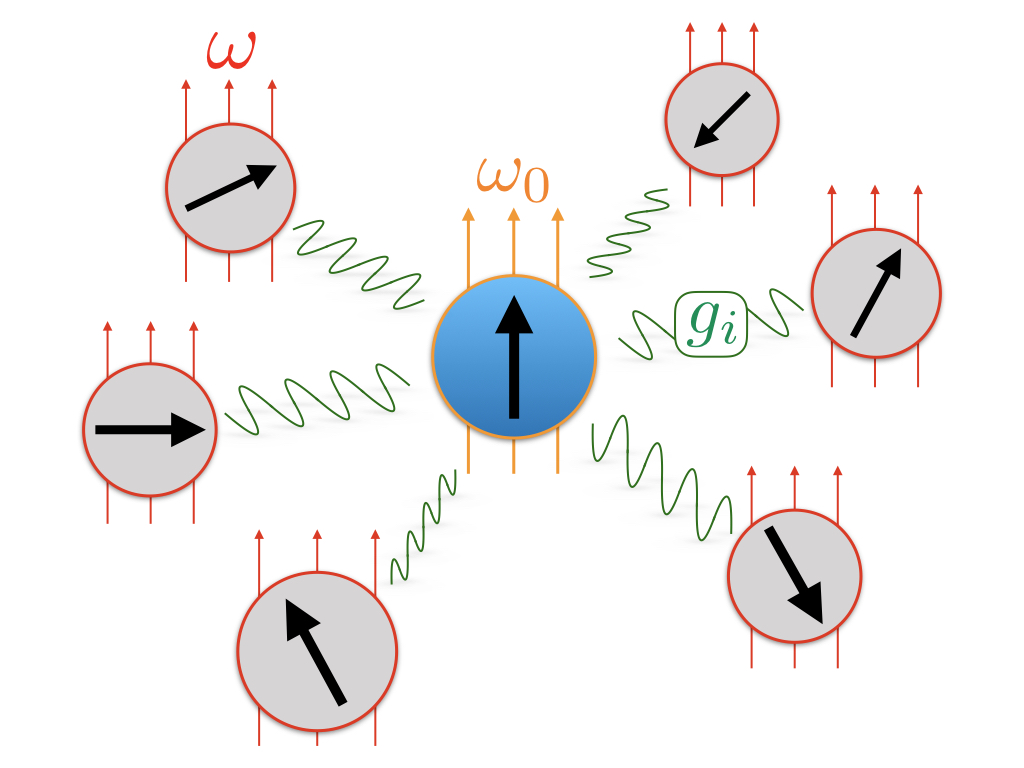}
\caption{ \textbf{Schematic of the spin-1/2 anisotropic central spin model}. A central qubit in a magnetic field of strength $\omega_0$ interacts with an environment of $L-1$ spin-1/2 particles in a uniform magnetic field of strength $\omega$ with interaction strengths $g_i$, $i = 1, 2, \dots, L-1$.\label{fig:model}}
\vspace{-\baselineskip}
\end{figure}

Central spin systems are typically described by a spin-1/2 Hamiltonian ($\hbar=1$):
\begin{equation}\label{eq:H}
H = \omega_0 \,S_0^z +   \omega \sum_{i=1}^{L-1} S_i^z + \sum_{i=1}^{L-1} \,g_i \left( S_0^x\, S_i^x + S_0^y\, S_i^y + \alpha\,S_0^z\, S_i^z \right),
\end{equation}
where $\omega_0$ is a local magnetic field on the central qubit, $\omega$ is a uniform magnetic field on the bath spins, $\alpha$ sets the anisotropy of the qubit-bath interaction, and $g_i$ sets the strength of the interaction between the central qubit and the $i$th bath spin for $i = 1, 2, \ldots, L-1$. Experimentally, the interaction strengths are inhomogeneous because of the randomness in the positions of the bath spins and/or the geometrical factors in dipolar interactions. For simplicity, we model the inhomogeneity as uncorrelated disorder, and take the $g_i$ to be independently and identically distributed uniformly in the interval $[1-\gamma, 1+\gamma]$ with $\gamma$ setting the disorder strength. Moreover, we study the model near resonance $\omega_0 = \omega - \alpha \sum_{j=0}
^{L-1} S_j^z$ where qubit-bath interactions are enhanced. Since $H$ conserves total spin magnetization $\sum_{j=0}^{L-1} S_j^z$, we set $\omega = 0$ without loss of generality.
Fig.~\ref{fig:model} shows a schematic of the model.

\begin{figure}[tp]
\centering
\includegraphics[width=\columnwidth]{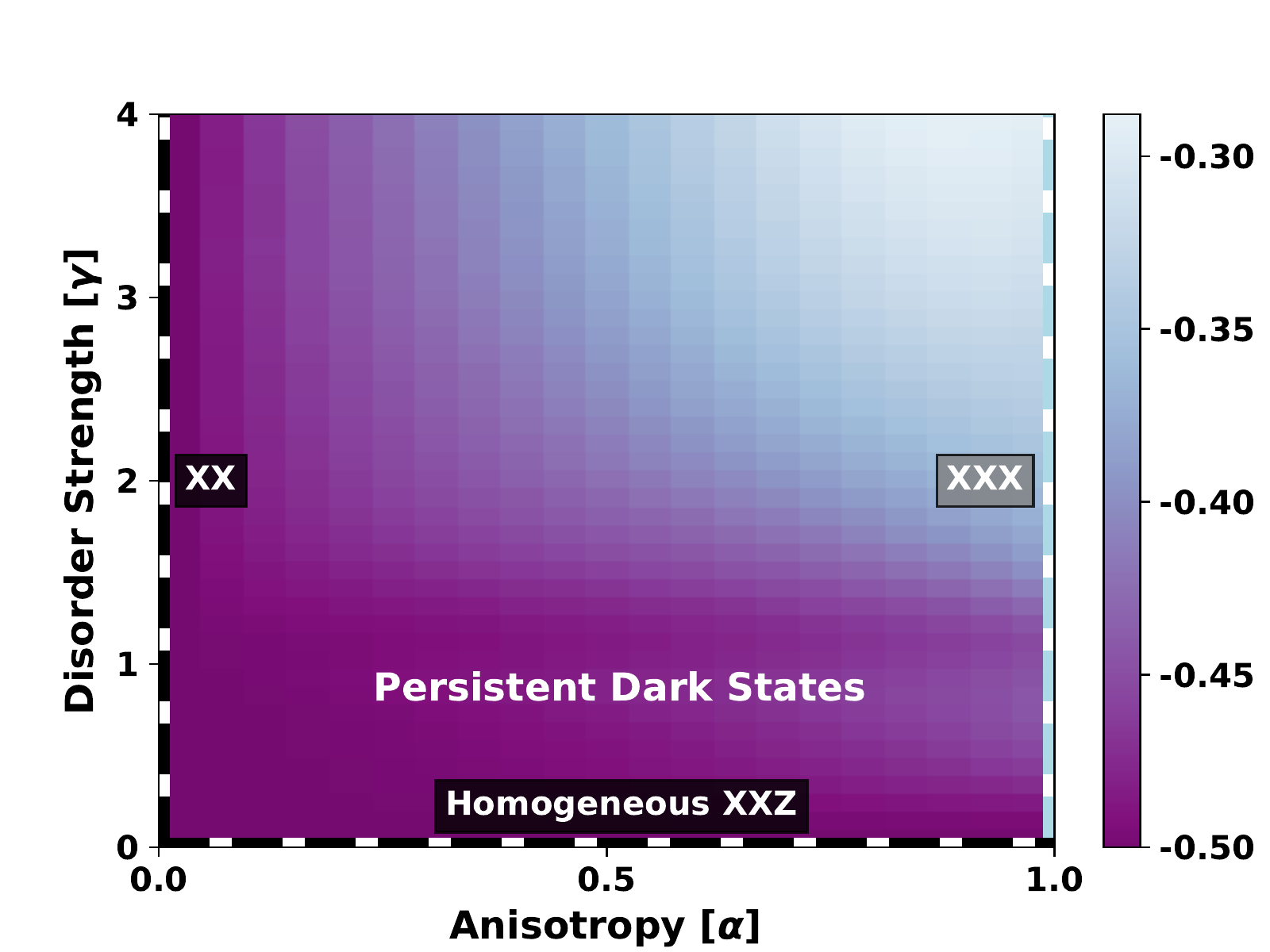}
\caption{ \textbf{Finite size crossover diagram.} Integrable lines with $N_D$ \exact dark eigenstates are shown as dashed black lines, while the integrable XXX line with no \exact dark eigenstates is shown in blue. In the chaotic regime between integrable lines, we show a color plot of the central spin polarization $[\overline{\langle S_0^z \rangle}]$, averaged over $N_{D}$ eigenstates with the smallest values of $|\langle S_0^z \rangle + 0.5|$ and $N_s = 400$ disorder samples in a fixed sector of total magnetization $\sum_j S_j^z=-1$ at resonance $\omega_0=\alpha$. In this regime, non-thermal \emph{persistent} dark states with $-0.5<[\overline{\langle S_0^z \rangle}] \ll 0$ (violet region) coexist with bright states whose central spin polarization is close to zero. 
While this crossover diagram shows a fixed system size $L=14$, we find no significant dependence on $L$ for system sizes amenable to numerical simulation. \label{fig:phase}}
\vspace{-\baselineskip}
\end{figure}

The Hamiltonian $H$ has three known integrable families. The first is the fully isotropic XXX model ($\alpha = 1$ and arbitrary $\gamma$), which describes systems with contact interactions such as quantum dots in $s$-type semiconductor bands~\cite{hanson_review_2007,schliemann2003electron}. This model belongs to the class of integrable XXX Richardson-Gaudin models~\cite{gaudin_bethe_2014,dukelsky_colloquium_2004,rombouts_quantum_2010}. The second is the fully anisotropic XX model ($\alpha = 0$ and arbitrary $\gamma$), which describes resonant exchange interactions in dipolar spin systems in rotating frames~\cite{hartmann1962nuclear,rovnyak2008tutorial,rao2019spin,lai2006knight}. It was only recently established that the XX model is integrable, arising as a singular limit of the class of hyperbolic XXZ Richardson-Gaudin models~\cite{villazon2020integrability}. 
The third is the homogeneous XXZ model ($\gamma = 0$ and arbitrary $\alpha$), which describes effective two-body interactions $H = \omega_0 \,S_0^z + g\,(S_0^x\, S^x + S_0^y\, S^y + \alpha\,S_0^z\, S^z) $ between the qubit and the collective spin of the bath $\vec{S} = \sum_{i=1}^{L-1} \vec{S}_i$~\cite{christ_nuclear_2009}. 
Fig.~\ref{fig:phase} shows the three integrable families in a broader phase diagram. 
The integrability of these models has enabled analytical and numerical studies of experimentally relevant systems using a variety of integrability-based techniques~\cite{schliemann2003electron,yuzbashyan_solution_2005,bortz_exact_2007,faribault_quantum_2009,schliemann_spins_2010,bortz_dynamics_2010,faribault_integrability-based_2013,claeys_spin_2017,nepomechie_spin-s_2018,he_exact_2019}.  

A remarkable feature of $H$ is that it exhibits dark eigenstates of a particularly simple product form when either $ \alpha = 0 \,$ or $ \,\gamma = 0 $~\cite{villazon2020integrability,christ_nuclear_2009}. These \exact dark states $\ket{\mathcal{D}}$ are states which exhibit no qubit-bath entanglement. Namely, they have a product state structure $\ket{\mathcal{D}}=\ket{\downarrow}_0 \otimes \ket{\mathcal{D}^-}$ or $\ket{\mathcal{D}}=\ket{\uparrow}_0 \otimes \ket{\mathcal{D}^+}$, in which the central spin is polarized along the $z$-direction, and the state of the bath satisfies
\begin{equation}\label{eq:DarkStates}
\left(\sum_{i=1}^{L-1} g_i S_i^{\pm} \right)\ket{\mathcal{D}^{\pm}} = 0.
\end{equation}
These states are furthermore independent of $\omega_0$ (and $\omega$), making the qubit state insensitive external axial fields, in addition to bath fluctuations. More generally, we define \emph{dark states} as states in which the qubit is nearly z-polarized and is not in thermal equilibrium with the surrounding bath. Dark states allow the surrounding spin bath to be used as a robust quantum memory~\cite{taylor_controlling_2003,taylor2003long,kurucz2009qubit,ding2014high}. Moreover, they pose limitations for dynamical nuclear polarization (DNP) experiments which attempt to polarize a mesoscopic bath by repeated qubit polarization and transfer~\cite{imamoglu_optical_2003,christ2007quantum,christ_nuclear_2009,belthangady2013dressed}. DNP protocols eventually prepare the system in a statistical mixture of dark states, producing effectively isolated qubits for decoherence-free quantum computation~\cite{lidar1998decoherence}.

In this work, we first establish that dark states are robust to integrability-breaking perturbations that tune the anisotropy $\alpha$ and the disorder strength $\gamma$. 
Specifically, in Sec.~\ref{sec:persistentdarkstates}, we show that dark states persist as exact eigenstates of the Hamiltonian, that only perturbatively mix with bright states (i.e. not dark states) over a broad range of values for $\alpha$ and $\gamma$, at system sizes amenable to numerical simulation. This perturbative mixing only slightly reduces the polarization of the qubit along the $z$-direction.
Remarkably, while $H$ is non-integrable/chaotic away from its integrable lines, dark states are well protected due to the presence of quasi-conserved charges (Sec.~\ref{sec:longliveddynamics}). 

To test the stability of dark states, we apply a recently developed exponentially sensitive probe for chaos, based on the scaling of the norm of the adiabatic gauge potential (AGP)~\cite{pandey2020adiabatic}. The AGP is defined as the operator which generates continuous adiabatic transformations between eigenstates and measures their sensitivity to perturbations of the underlying Hamiltonian~\cite{Kolodrubetz,Demirplak1,Berry,villazon_heat_2019,Bukov1}. Its norm is closely related to the quantum geometric tensor and the fidelity susceptibility~\cite{Kolodrubetz, Venuti_2007, Sierant_2019}. The norm of the AGP was found to scale exponentially with system size for chaotic perturbations, in contrast to integrable perturbations leading to polynomial scaling~\cite{pandey2020adiabatic}. 

In our present context, the AGP norm grows exponentially in accordance with quantum chaos (Sec.~\ref{sec:chaos}), but interestingly, the growth rate of the logarithm of this norm is twice the rate expected for ergodic systems satisfying the eigenstate thermalization hypothesis (see Fig.~\ref{fig:AGPnorm} in Sec.~\ref{sec:chaos}).
This rate saturates the upper bound for eigenstate sensitivity to perturbations~\cite{pandey2020adiabatic}. It reflects a very strong mixing between neighbouring eigenstates of the system and leads to ultra-slow (exponentially long in system size) relaxation dynamics (Sec.~\ref{sec:relaxation}), reminiscent of the Arnold diffusion in classical near-integrable systems \cite{arnold_mathematical_1989}. While a similar behavior of the AGP norm was found in the previous work~\cite{pandey2020adiabatic} for spin chains with weak integrability breaking perturbations, here we find that this chaotic but non-ergodic (CNE) regime extends to large perturbation strengths, even for the largest system size that we are able to simulate. Sec.~\ref{sec:discussion} is reserved for discussion and conclusions.

\section{Persistent Dark States}
\label{sec:persistentdarkstates}

Away from the integrable lines in Fig.~\ref{fig:phase}, the eigenstates $\ket{n}$ of $H$ no longer admit exact \exact dark states. Nevertheless, we can identify \emph{persistent} dark eigenstates with approximate product form using (i) the eigenstate expectation value of the central qubit $z$-projection $\langle S_0^z \rangle \equiv \langle n | S_0^z | n \rangle$, or (ii) the eigenstate entanglement entropy $\mathcal{S}_{E}^{0}$ of the qubit. The latter is defined as 
\begin{equation}
\mathcal{S}_{E}^{0} \equiv -\Tr( \rho_{0} \,\log(\rho_{0}) ) , \quad\quad \rho_{0} \equiv \Tr_{\mathcal{B}}(\rho),
\end{equation}
where $\rho_0$ is the reduced density matrix for the qubit obtained by tracing out the bath $\mathcal{B}$ degrees of freedom, and $\rho = \ket{n} \bra{n}$ is the density matrix of eigenstate $\ket{n}$ with energy $E_n$.

Consider for reference the XX model ($\alpha=0$) at resonance ($\omega_0 = 0$) in a sector with negative net magnetization ($\sum_{j=0}^{L-1} S_j^z < 0$)~\cite{villazon2020integrability}. For any \exact dark eigenstate $|\mathcal{D}\rangle$ of this model, $\langle \mathcal{D} | S_0^z | \mathcal{D} \rangle = -1/2 $ and $\mathcal{S}_E^0 = 0$. On the other hand the bright eigenstates $|\mathcal{B}\rangle$ of the model satisfy $\langle \mathcal{B} | S_0^z | \mathcal{B} \rangle = 0$ and $\mathcal{S}_E^0 = \ln(2)$ at resonance\footnote{Far from resonance, the central spin is nearly polarized even in the bright states of the XX model: $\langle \mathcal{B} | S_0^z | \mathcal{B} \rangle \approx \pm 1/2 $ and $ \mathcal{S}_E^0 \approx 0$.}. 

For $\alpha>0$, we find eigenstates $| \mathcal{D}(\alpha) \rangle$ of $H$ that are adiabatically connected to $| \mathcal{D}(0) \rangle \equiv | \mathcal{D} \rangle$ by a unitary transformation generated by the adiabatic gauge potential (AGP) $\mathcal{A}_{\alpha}$:
\begin{equation}\label{eq:persistent-alpha}
| \mathcal{D}(\alpha) \rangle = \exp\left( -\,i \int_{0}^{\alpha} \mathcal{A}_{\alpha'} \,d\alpha' \right) | \mathcal{D} (0)\rangle.
\end{equation}
In these eigenstates, both the z-projection and entanglement entropy of the qubit will deviate from their $\alpha=0$ values. The question becomes whether these deviations are perturbatively small in $\alpha$, and how this depends on the system size $L$. In chaotic systems, the AGP is generally a highly non-local many-body operator with an exponentially large norm~\cite{Kolodrubetz,sels2017minimizing,pandey2020adiabatic}. In the present context, the parameter $\alpha$ breaks the integrability of the system (see Sec.~\ref{sec:chaos}). Naively, we expect qubit observables in $| \mathcal{D}(\alpha) \rangle$ to perturbatively connect to their values in $| \mathcal{D}(0) \rangle$ only for $\alpha$ that is exponentially small in the system size. 

\begin{figure}[htp]
\includegraphics[width=\columnwidth]{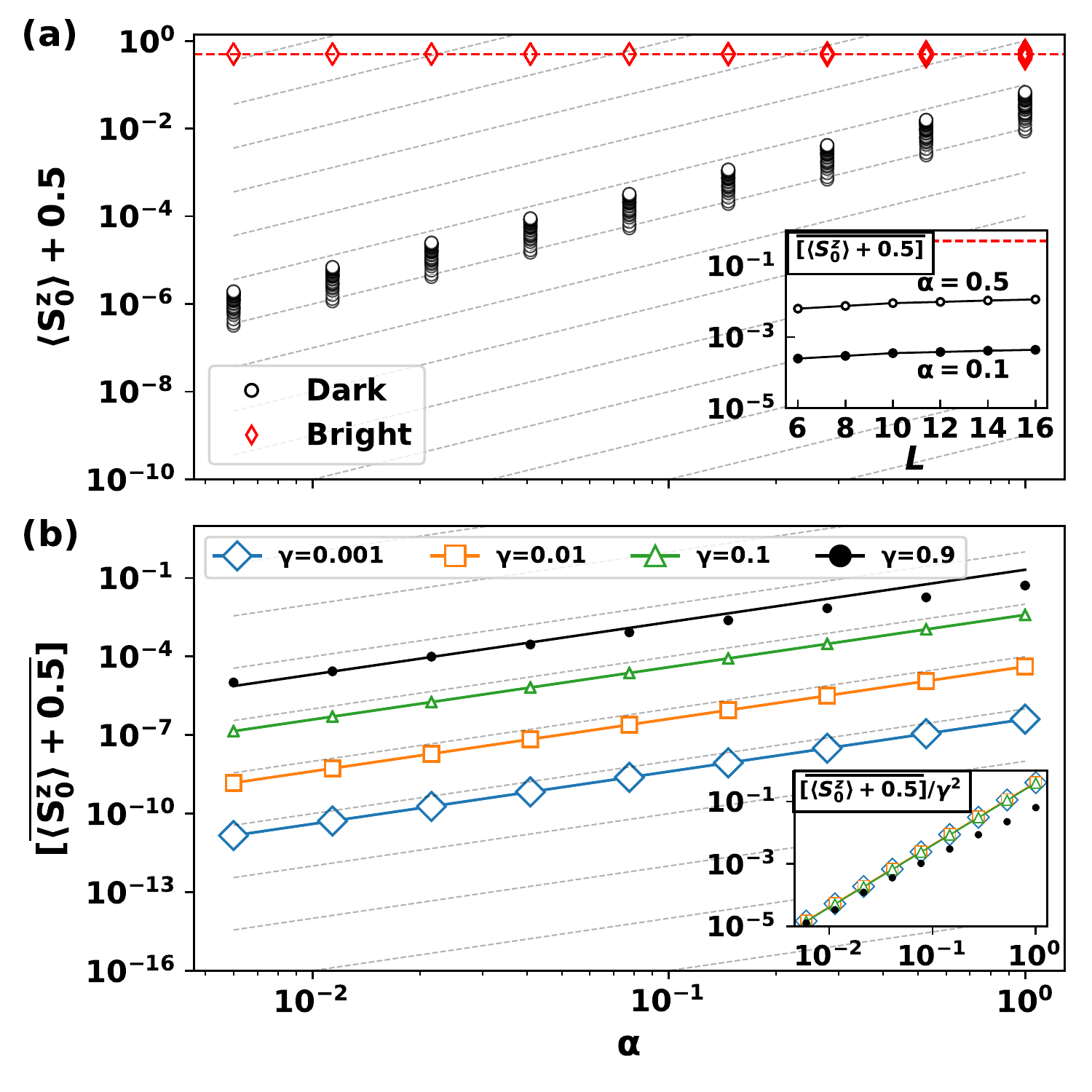}
\caption{ \textbf{Dark states persist away from the integrable lines at finite size.} (a) Upper panel: Expectation value of the central spin $z$-projection for every eigenstate of $H$ in a typical sample as a function of $\alpha$. Persistent dark (black circles) and bright (red diamonds) states are easily distinguished by their value of $\langle S_0^z \rangle$. Dotted lines (gray) show $\alpha^2$ scaling, while the horizontal dashed line (red) indicates $\langle S_0^z \rangle=0$. Inset: System size dependence of $\langle S_0^z \rangle$ averaged over $N_s$ disorder samples and the $N_D$ eigenstates with smallest z-projection: $\overline{[\langle S_0^z \rangle + 0.5]}$. (b) Lower panel: $\overline{[\langle S_0^z \rangle + 0.5]}$ (markers) as a function of $\alpha$ for several values of $\gamma$. The solid lines plot the perturbative prediction of Eq.~\eqref{eq:pt-alpha}. Inset: Upon re-scaling the vertical axis by $\gamma^2$, the curves collapse onto a single curve. Parameters: $L = 12$, $\omega_0=\alpha$, $\sum_{j} S_j^z=-1$, $N_s=500$, and in (a) $\gamma = 0.5$. \label{fig:pt}}
\vspace{-\baselineskip}
\end{figure}

Remarkably, at numerically accessible system sizes, we find that $\mathcal{A}_{0}\equiv \mathcal \,\mathcal{A}_{\alpha} (\alpha\to0^{+})$ can be well-approximated by few-body operators and that perturbation theory works exceedingly well to characterize qubit observables in $|\mathcal{D}(\alpha)\rangle$. To illustrate, consider the perturbative expansion of the $S_0^z$ expectation to leading order in $\alpha$:
\begin{align} \nonumber 
\langle \mathcal{D}(\alpha) | \,S_0^z\, | \mathcal{D}(\alpha) \rangle &- \langle \mathcal{D}(0) | \,S_0^z\, | \mathcal{D}(0) \rangle \\ 
= \frac{\alpha^2 }{2} & \langle \mathcal{D}(0)|\, [\mathcal{A}_0,[S_0^z, \mathcal{A}_0]] \,|\mathcal{D}(0) \rangle + \cdots \label{eq:pt-alpha}
\end{align} 
The leading term is of order $\alpha^2$, as the coefficient of the linear in $\alpha$ term, $\langle \mathcal{D}(0) |i [\mathcal{A}_0,S_0^z]| \mathcal{D}(0)\rangle = 0$, vanishes because $S_0^z |\mathcal{D}(0)\rangle = \pm 1/2 |\mathcal{D}(0)\rangle$. Fig.~\ref{fig:pt} numerically demonstrates that the left-hand side $\langle S_0^z \rangle + 0.5$ scales as $\alpha^2$ for a subset of the eigenstates over several orders of magnitude of $\alpha$ and $\gamma$. 

In more detail, when $\alpha >0$, the resonance condition in a given polarization sector is shifted by the mean anisotropy to $\omega_0 = - \alpha \sum_{j=0}^{L-1} S_j^z$ (see Supplemental Information). For concreteness, we focus on a single polarization sector $\sum_{j=0}^{L-1} S_j^z = -1$, such that the resonance occurs at $\omega_0 = \alpha$. Fig.~\ref{fig:pt}(a) shows numerical computations of the expectation value $\langle S_0^z \rangle + 0.5$ in every eigenstate of $H$ at moderate disorder strength ($\gamma=0.5$) over several orders of magnitude in $\alpha$. Persistent dark states (black/dark circles) are easily identifiable, as they connect smoothly to $\langle S_0^z \rangle \to -0.5$ as $\alpha\to0$. The deviation from $-0.5$ scales as $\sim \alpha^2$, consistent with Eq.~\eqref{eq:pt-alpha} (dotted lines). The bright states (red/light diamonds) are similarly perturbed around their value at resonance $\langle S_0^z \rangle = 0$ (dashed red horizontal line). As $\alpha\to1$, dark and bright states attain comparable central spin projections. The inset of Fig.~\ref{fig:pt}(a) shows the system-size dependence of the averaged expectation value $[\overline{\langle S_0^z \rangle+0.5}]$, where $\overline{\,\,.\,\,}$ denotes an average over all $N_D$ persistent dark states, and $[\,\,.\,\,]$ denotes an average over $N_{s}$ disorder samples. In eigenstates that satisfy the ETH, the expectation value $\langle S_0^z\rangle_{\mathrm{th}} = \sum_j S
^z_j/L$ approaches zero with increasing $L$, in a sector with fixed magnetization. However, we find that $\langle S_0^z\rangle$ approaches its thermal value only very slowly with system size $L$, suggesting that dark state properties persist to system sizes much larger than we probe here. From the current analysis we cannot conclude whether or not they survive the thermodynamic limit.

Fig.~\ref{fig:pt}(b) shows $[\overline{\langle S_0^z \rangle+0.5}]$ for varying disorder strengths $\gamma$. The markers show numerical data and the solid lines show the analytic predictions given by Eq.~\eqref{eq:pt-alpha} up to $\mathcal{O}(\alpha^2)$. Again, we find leading order perturbation theory to  be in excellent agreement with numerical simulations for $\gamma < 1$ and the entire $\alpha$ range between the integrable points $\alpha = 0$ and $\alpha =1$. As $\gamma \to 1$ perturbation theory begins to break down (see $\gamma = 0.9$ line in plot). When $\gamma \gtrsim 1$, perturbation theory breaks down much faster at $\alpha \ll 1$ (see Supplemental Information). 

Persistent dark states are well captured by perturbation theory due to the quasi-locality of $\mathcal{A}_{0}$ at numerically accessible system sizes. To see this, we decompose the AGP into $k$-body operators:
\begin{equation}
\mathcal{A}_{0} = \sum_{k=1}^{L} \sum_{\{p_i\}}\sum_{\{\lambda_j\}} J^{p_1, \ldots ,p_k}_{\lambda_{1}, \ldots, \lambda_{k}} \,  \sigma^{\lambda_1}_{p_1} \cdots \sigma^{\lambda_{k}}_{p_k}
\end{equation}
Here $\sigma^{\lambda_j}_{p_i}$ with $\lambda_j \in \{x,y,z\}$ denote the Pauli basis operators on site $p_i$, where $0\leq p_1 < p_2 < \ldots < p_k \leq L-1$ for every $k = 1,\ldots,L$. Throughout this work, we define the norm of any operator $\Theta$ by its normalized Frobenius norm: 
\begin{equation}
\| \Theta \|^2 = \frac{1}{2^{L}}  \mathrm{Tr}\big(\Theta^{\dagger}\,\Theta\big).
\end{equation}
We find $\mathcal{A}_{0}$ has non-zero weight only for $k$-body operators with $k=3,5,7,\dots$. Moreover, the total weight of k-body operators decays as $1/k^{c}$ for $c>0$, so that $\mathcal{A}_{0}$ is well-approximated by 3-body operators. In the Supplemental Information, we showcase the quasi-locality of $\mathcal{A}_{0}$ and estimate $c\approx 3$.

One can similarly find persistent dark states based on translations in the $\gamma$-parameter space, as in Eq.~\eqref{eq:persistent-alpha}, with a different adiabatic gauge potential $\mathcal{A}_{\gamma}$. We find perturbatively accessible persistent dark states away from the $\gamma = 0$ line at numerically accessible system sizes. The inset of Fig.~\ref{fig:pt}(b) shows the re-scaled averaged expectation value $[\overline{\langle S_0^z \rangle + 0.5}]/\gamma^2$ vs $\alpha $ at resonance. The data collapse for $\gamma =0.001, 0.01, 0.1$ shows that:
\begin{equation}
[\overline{\langle \mathcal{D}(\alpha,\gamma)| S_0^z | \mathcal{D}(\alpha,\gamma) \rangle + 0.5}] \propto \gamma^2 \alpha^2
\end{equation}
at small $\gamma, \alpha$, in perfect agreement with the perturbative result. At larger values of $\gamma$ ($\gamma=0.9$), we see deviations from the perturbative result as $\alpha \to 1$. Persistent dark states were previously found by mapping exact \exact dark eigenstates from the homogeneous isotropic limit $(\alpha,\gamma) = (1,0)$ to the inhomogeneous isotropic regime $\alpha =1, \gamma > 0$ \cite{christ2007quantum, taylor_controlling_2003}. Our results perturbatively extend dark states into a broader region of parameter space at finite size (see Fig.~\ref{fig:phase}).

\begin{figure}[htp]
\includegraphics[width=\columnwidth]{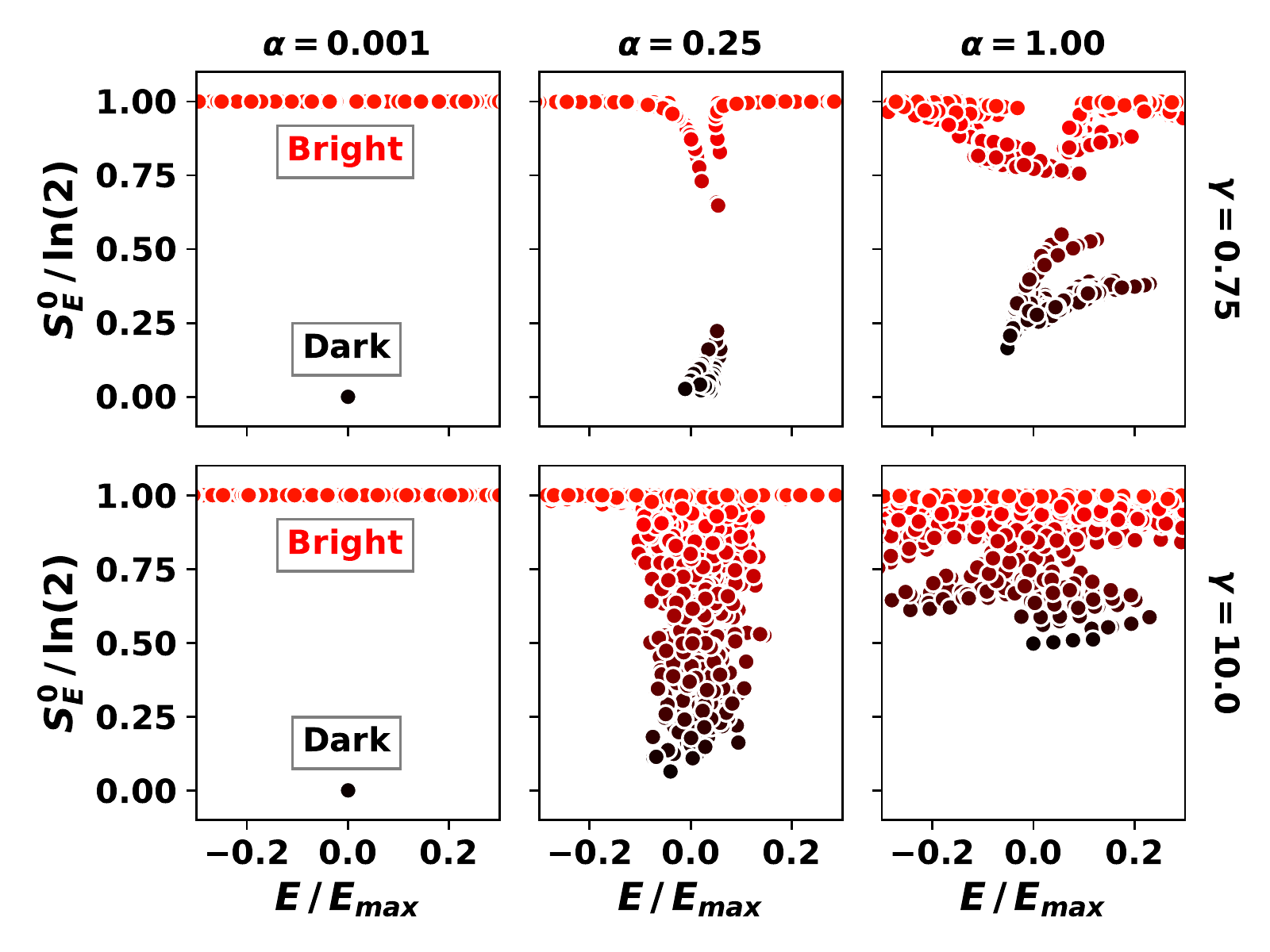}
\caption{ \textbf{Low central spin entanglement entropy reveals persistent dark states.} The entanglement entropy of the central spin in eigenstates vs the re-scaled energy in typical samples at low (top panels) and large (bottom panels) values of $\gamma$. In the top panels, the persistent dark states have low central-spin entanglement at all $\alpha$ values. At larger disorder (bottom panels), the entropy approaches $\ln(2)$ for all eigenstates as $\alpha\to1$. Parameters:  $L=16$, $\sum_{j} S_j^z = -1$,  $\omega_0 =  \alpha$, $E_\mathrm{max}$ is the maximum energy of $H$ in the given total magnetization sector.}
\label{fig:EE}
\vspace{-0\baselineskip}
\end{figure}

Persistent dark states can also be identified by their low central spin entanglement entropy (see Fig.~\ref{fig:EE}). For moderate to low disorder (upper panels with $\gamma=0.75$), low entanglement (dark) states persist and do not fully mix with high entanglement (bright) states, even as $\alpha \to 1$. At sufficiently large disorder (lower panels with $\gamma=10.0$), the persistent dark state picture breaks down as $\alpha \to 1$, since most states acquire large central spin entanglement $S_E^0\approx \ln(2)$.

\section{Quasi-Conserved Operators} 
\label{sec:longliveddynamics}

The question of whether and how systems thermalize is a fundamental one in quantum statistical mechanics. Steady states of integrable systems typically have non-thermal correlations due to the presence of extensively many conserved quantities, and are described by Generalized Gibbs Ensembles (GGEs) that account for these conserved quantities~\cite{jaynes1957information,rigol2007relaxation,vidmar2016generalized}. Generic integrability-breaking perturbations usually yield Hamiltonians which are chaotic and satisfy the Eigenstate Thermalization Hypothesis (ETH)~\cite{d2016quantum,deutsch2018eigenstate}. Nevertheless, the integrable Hamiltonian can control the approach to a long-lived pre-thermal state when the strength of the integrability-breaking perturbation is sufficiently small~\cite{mori2018thermalization}. 

In this section, we establish that the central spin model in Eq.~(\ref{eq:H}) has long-lived non-thermal states controlled by the XX and XXZ integrable lines at accessible system sizes. Specifically, we show that $H$ has approximate conservation laws that persist away from the integrable lines, giving rise to non-thermal correlations in local qubit observables. A simple way to detect the non-thermal correlations in quench experiments is through observables, such as $\langle S_0^z \rangle$, that differentiate between dark and bright states. 
As $\langle S_0^z \rangle$ takes non-thermal values (close to $\pm 1/2$) in the dark state manifold, we find quenched steady states that retain memory of the initial $z$-polarization of the central spin. 

The integrable lines of the model ($\alpha=0$, $\alpha=1$, and $\gamma=0$) constitute families of Richardson-Gaudin models with extensive numbers of \emph{bilinear} two-body conserved charges $Q_i$~\cite{villazon2020integrability,claeys2018richardson}. Upon breaking integrability, there no longer exists an extensive number of exactly conserved charges. Instead we find an extensive number of quasi-conserved charges, which very nearly commute with $H$. To find such quasi-conserved charges, we numerically construct an exhaustive set of few-body operators $Q_k$. These operators are conserved iff $\|[H,Q_k]\|=0$. The quasi-conserved charges are those operators $Q_k$ with very small ratio:  $\|\,[H,Q_k]\,\|/\|Q_k\|\ll \Gamma_{\mathrm{typ}}$, where $\Gamma_{\mathrm{typ}} \equiv \sqrt{\,\| H \|^2/L\,}$ sets a typical energy scale.

We construct $Q_k$ using the ansatz:
\begin{align}
    Q_k = \sum_i q_i \,\theta_i \label{Qexp}
\end{align}
and restrict $\{\theta_i\}$ to a complete set of $m$ trace-orthogonal one and two-body spin-1/2 operators with unit norm. To avoid cumbersome notation, we leave the dependence of $q_i$ and $\theta_i$ on $k$ implicit. 
We further set $\|Q_k \|^2 = \sum_j |q_j|^2 = 1$. To determine the coefficients $q_j$, we solve the eigenvalue problem:
\begin{equation}\label{eq:eig}
M \,\vec{q} = \Gamma^2 \,\vec{q}, \quad\quad \vec{q} \equiv (q_1,q_2,\dots,q_{m})
\end{equation}
where $M_{ij} \equiv \mathrm{Tr}([H,\theta_i][H,\theta_j])/2^{L}$ and we take $\Gamma>0$. The eigenvectors of $M$ then yield through Eq.~\eqref{Qexp} a set of orthogonal and bilinear operators with known decay properties. Specifically, the eigenvalue $\Gamma^2$ equals the norm of the commutator:
\begin{equation}
\|\,[H,Q_k]\,\|^2 = \sum_i\sum_j q_i^{*} M_{ij} q_j = \Gamma^2.
\end{equation}
To connect $\Gamma$ to the operator decay rate, consider the short-time expansion of the symmetrized unequal time correlator of $Q_k$ at infinite temperature~\cite{Kim_2015, Sho_2020}:
\begin{align}
    \frac{1}{2^{L}}\Tr\left( \frac{Q_k^\dagger(t)Q_k(0) + Q_k^\dagger(0) Q_k(t)}{2} \right) = 1 &- \frac{t^2}{2} \|[H,Q_k]\|^2 \nonumber \\
    &+ \mathcal{O}(t^4).
\end{align}
The correlator's decay rate is thus given by $\Gamma = \|[H,Q_k]\|$. Hence, if $\Gamma=0$, the unequal-time correlator equals one for all $t$. If instead $0<\Gamma \ll \Gamma_{\mathrm{typ}}$, where $\Gamma_{\mathrm{typ}}$ sets a typical decay rate, then the correlator is close to one for a long time $1/\Gamma$ and $Q_k$ is approximately conserved up to this time. 

Away from the integrable lines, we generally find three kinds of eigenvalues $\Gamma^2$: a few $\mathcal{O}(1)$ zero eigenvalues, $\mathcal{O}(L^2)$ large $\Gamma^2 \approx \Gamma^2_{\mathrm{typ}}$ eigenvalues, and an extensive number $\mathcal{O}(L)$ of eigenvalues with $\Gamma^2 \ll \Gamma^2_{\mathrm{typ}}$. Zero eigenvalues correspond to exactly conserved charges related to known conservation laws, while the extensive number of small positive eigenvalues can be identified with quasi-conserved charges.

\begin{figure}[tp]
\includegraphics[width=\columnwidth]{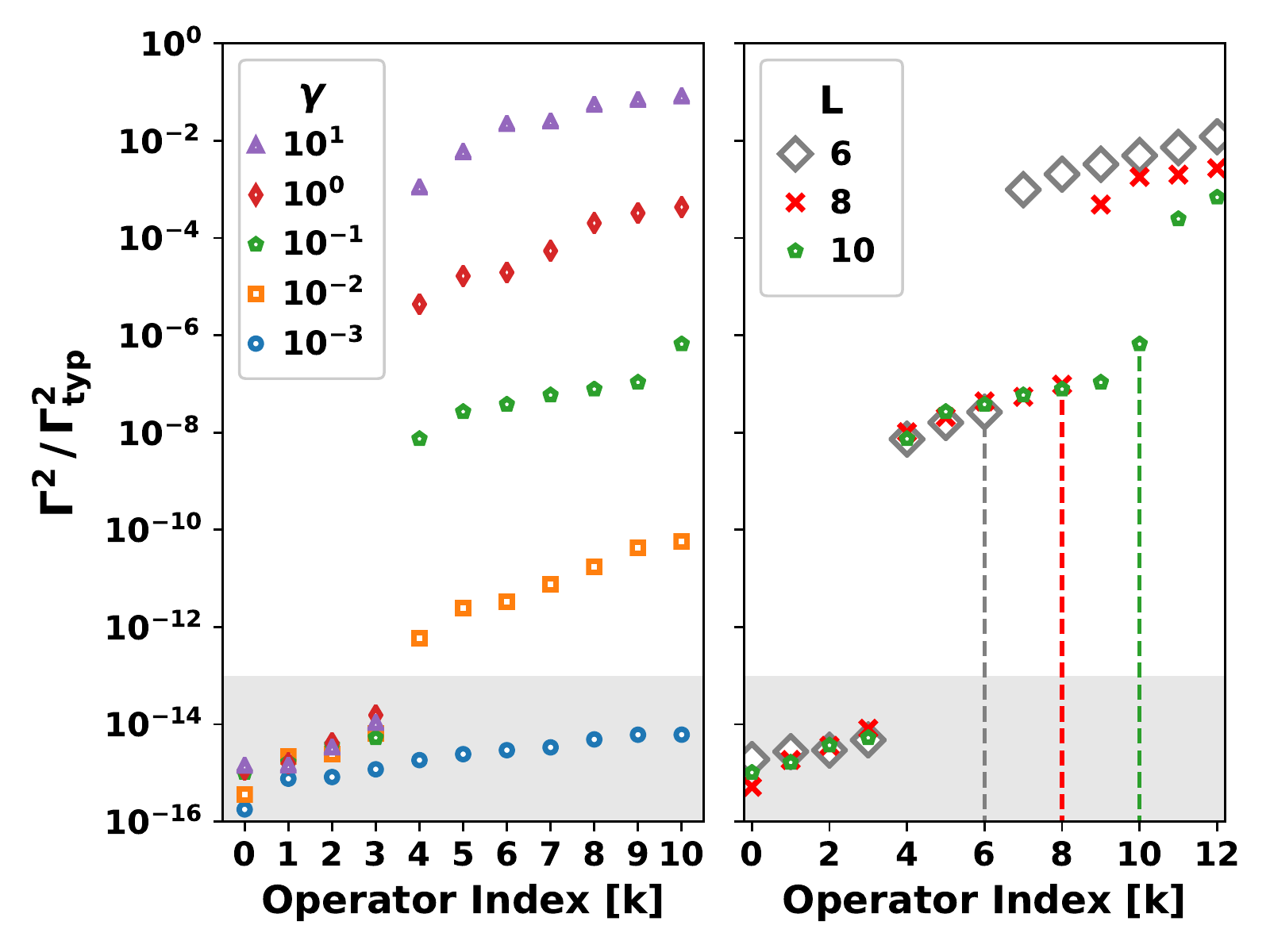}
\caption{ \textbf{An extensive number of quasi-conserved quadratic charges persist upon breaking integrability.} The smallest eigenvalues of Eq.~\eqref{eq:eig} in dimensionless units at fixed $L=10$ for a typical sample at different disorder strengths $\gamma$ (left) and at different $L$ for fixed $\gamma=0.1$ (right). Values in the shaded region are zero within numerical accuracy. The left panel shows that the dimensionless decay rate $\Gamma/\Gamma_{\mathrm{typ}}$ of the quasi-conserved operators increases with $\gamma$, while the right panel shows that the number of quasi-conserved operators is extensive $\sim L$ (vertical lines denote the largest quasi-conserved index for each $L$). Parameters: $\omega_0=\alpha=0.5$, (left) $L=10$, (right) $\gamma=0.1$.
\label{fig:Qs}}
\vspace{-\baselineskip}
\end{figure}

Fig.~\ref{fig:Qs} shows the smallest eigenvalues $\Gamma^2$ (re-scaled by $\Gamma_{\mathrm{typ}}^2$) obtained by numerically solving Eq.~\eqref{eq:eig} as a function of the index $k$ of the corresponding operator $Q_k$. The right panel shows the eigenvalues at several system sizes ($L=6,8,10$) for a fixed disorder strength $\gamma=0.1$. We find $4$ eigenvalues which are zero within numerical accuracy (shaded gray region), corresponding to exactly conserved charges; namely $H$, $\sum_{j} S_j^z$, $H^2$, $(\sum_{j} S_j^z)^2$. We also see a cluster of intermediate eigenvalues corresponding to quasi-conserved charges, which are separated by a gap from a set of larger eigenvalues (only a small fraction of this set is shown). The vertical dashed lines mark the indices of the quasi-conserved operators $Q_k$ with largest eigenvalue for each system size. These maximal indices increase in precise proportion to $L$, showing that the number of quasi-conserved charges is extensive. Furthermore, the eigenvalues themselves remain of the same order of magnitude for the different system sizes, indicating that the operator lifetimes have no significant system size dependence up to $L=10$. 

The left panel of Fig.~\ref{fig:Qs} shows the $(L+1)$ smallest eigenvalues of Eq.~\eqref{eq:eig} for several values of disorder strength $\gamma$ at $L=10$. As expected, we find exactly $(L+1)$ conserved charges as we approach the integrable line $\gamma = 0$ (see $\gamma = 0.001$ data within gray region). On increasing $\gamma$, only $4$ charges remain exactly conserved, while the remaining $(L-3)$ charges become quasi-conserved. The lifetime ($\propto 1/\Gamma$) of the quasi-conserved charges furthermore systematically decreases with increasing $\gamma$. Previous studies have found similar long-lived quasi-conserved charges in a family of near-(Richardson-Gaudin)-integrable spin models with all-to-all interactions~\cite{bentsen2019integrable}. 

\begin{figure}[tp]
\includegraphics[width=\columnwidth]{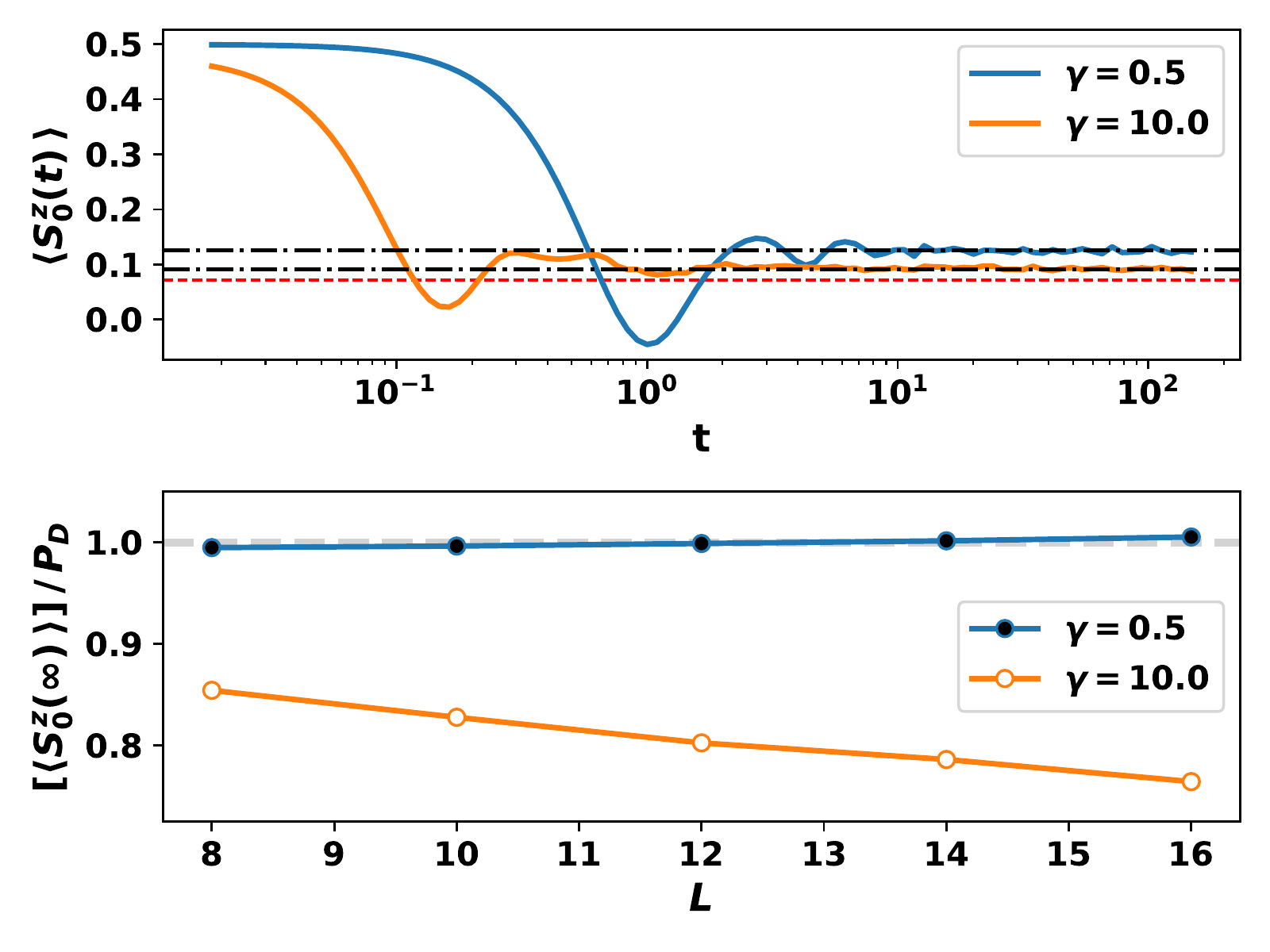}
\caption{ \textbf{Central spin $z$-polarization retains memory in the pre-thermal state.} Top: time evolution of $\langle S_0^z(t) \rangle$ for a quench to resonance ($\omega_0 = -\alpha$) from the initial density matrix $\rho_i$ (Eq.~\eqref{Eq:rhoidef}) in a typical sample at two different disorder strengths. The late time values (horizontal black dash-dotted lines) differ from the thermal value (horizontal dashed red line). Bottom: The ratio of the disorder-averaged long-time value $[\langle S_0^z(\infty) \rangle]$ to the long-time value on the integrable line $P_D$ vs $L$. The ratio is close to one with no significant finite-size flow at moderate $\gamma=0.5$, but decreases with increasing $L$ at large $\gamma=10.0$. Parameters: $L=14$ (top), $N_s=1000$ (bottom), $\alpha=0.75$, $\sum_j S_j^z=+1$. \label{fig:long-time}}
\vspace{-\baselineskip}
\end{figure}

The extensively many two-body quasi-conserved charges $Q_k$ control long-lived non-thermal states in quench experiments. The top panel of Fig.~\ref{fig:long-time} shows the relaxation of $\langle S_0^z(t) \rangle$ to a non-thermal value in a typical sample at moderate and large disorder strength. The system is initialized in a polarization sector $\sum_{j=0}^{L-1} S_j^z = + 1$ far from resonance ($\omega_0 = 50$) in the mixed state
\begin{align}
    \rho_i = \ket{\uparrow}_0 \bra{\uparrow}_0 \otimes \mathbbm{1} \label{Eq:rhoidef},
\end{align}
with the bath spins at infinite temperature and the central spin maximally polarized along $+z$. The top panel plots $\langle S_0^z(t) \rangle \equiv \Tr (\rho_i S_0^z(t))/\Tr(\rho_i)$ following a quench to resonance ($\omega_0 = -\alpha$). We observe a fast decay to a positive value that is different from the thermal value $\langle S_0^z \rangle_{\mathrm{th}}= \sum_j S_j^z /L = 1/L$. Thus, $\langle S_0^z(\infty) \rangle$ retains memory of its initial condition at these system sizes. This memory is a consequence of the weight of $\rho_i$ on the persistent dark state manifold. 

Two comments are in order. First, the hybridization between the dark and bright state manifolds increases with disorder strength. Consequently, at any given $\alpha$, $|\langle S_0^z \rangle|$ in the persistent dark state manifold decreases with increasing $\gamma$ (see Fig.~\ref{fig:pt}). This explains why $\langle S_0^z (\infty) \rangle$ decreases with increasing $\gamma$ in Fig.~\ref{fig:long-time}. Next, the initial decay in the top panel of Fig.~\ref{fig:long-time} is a consequence of dephasing between the perturbed bright states. As $\langle S_0^z \rangle \approx 0$ in each perturbed bright eigenstate, the weight of $\rho_i$ on the perturbed bright states does not contribute to the non-zero value of $\langle S_0^z(\infty) \rangle$. 

The lower panel of Fig.~\ref{fig:long-time} plots the re-scaled and disorder-averaged long time value $[\langle S_0^z(\infty) \rangle]$ with $L$. The re-scaling factor $P_{D}$ is the expected polarization of the central spin due to the weight of $\rho_i$ on the dark manifold,
\begin{align}
P_D \equiv \frac{1}{2} \frac{N^{\uparrow}_{D}}{N^{\uparrow}_{D} + N^{\uparrow}_{B}}
\end{align}
where $N^{\uparrow}_{D}$ and $N^{\uparrow}_{B}$ are respectively the number of dark and bright states with the central spin pointing along $+z$ in the appropriate polarization sector at large $\omega_0$. On the integrable XX line at $\alpha=0$, we expect that $[\langle S_0^z(\infty) \rangle] = P_{D}$. The blue (filled) curve in the lower panel of Fig.~\ref{fig:long-time} is perturbatively accessible from the integrable line at the numerically accessible system sizes, and thus we find that $[\langle S_0^z(\infty) \rangle]/P_{D}$ is close to one. At larger disorder strength however, the hybridization between the dark and bright states increases with $L$ at the accessible sizes. The long time value $[\langle S_0^z(\infty) \rangle]/P_D$ is thus smaller than the long time value at $\alpha=0$, with the discrepancy growing with $L$ (see orange curve with open markers). At large disorder strength, $[\langle S_0^z(\infty) \rangle]/P_D$ shows a trend toward the thermal value $\langle S_0^z \rangle_{\mathrm{th}}/P_D = 0.5$ with increasing $L$. This thermal value follows from $\langle S_0^z \rangle_{\mathrm{th}} = 1/L$ and $P_D \sim 2/L$ as $L\to\infty$ in the sector with $\sum_j S_j^z= +1$ total magnetization\footnote{Given a total magnetization density $s\equiv \sum_j S_j^z/L$, $\langle S_0^z \rangle_{\mathrm{th}}/P_D \to \textrm{min}(0.5-s, 0.5+s)$ as $L \to \infty$. The ratio is less than one for all $s \in (-0.5,0.5)$.}. At the given system sizes, we cannot determine with certainty whether the qubit will saturate at or before it reaches its thermal value, even in the presence of strong disorder. 

We conclude this section with two remarks. First, in addition to $S_0^z$, generic two-body observables with significant overlap on the quasi-conserved charges are expected to exhibit similar non-thermalizing behavior and non-thermal eigenstate expectation values. While it is possible that in the thermodynamic limit $L\to\infty$ all such observables will thermalize, there is no indication that this will happen from available data at small or intermediate disorder. Even if it happens, the non-thermal state after the quench could crossover to an extremely long-lived and very stable prethermal regime.

\section{Chaotic but non-ergodic regime}
\label{sec:chaos}
In previous sections, we established that dark eigenstates persist on adding putative integrability-breaking perturbations to Eq.~\eqref{eq:H} at numerically accessible system sizes. At these sizes, the model thus does not satisfy the ETH and few-body observables do not thermalize in isolation. However, we expect the eigenstate and dynamical behavior to change with increasing $L$. In this section, we provide evidence that the model is in a chaotic non-ergodic regime (CNE) characterized by an exponential sensitivity of eigenstates to small perturbations and the presence of relaxation times that are exponentially long in the system size.

Energy level statistics are a widely used tool to diagnose chaos and predict thermalization~\cite{poilblanc1993poisson,casati1985energy,atas2013distribution}. Integrable systems generally follow Poisson level statistics, while chaotic systems exhibit Wigner-Dyson statistics due to level repulsion in accordance with random matrix theory~\cite{d2016quantum}.  The use of level statistics to diagnose chaos is limited to relatively small system sizes where exact diagonalization can be reasonably implemented. For our present model, level statistics show weak-to-negligible level repulsion, thus proving insufficient to establish chaos.  

\begin{figure}[bp]
\includegraphics[width=\columnwidth]{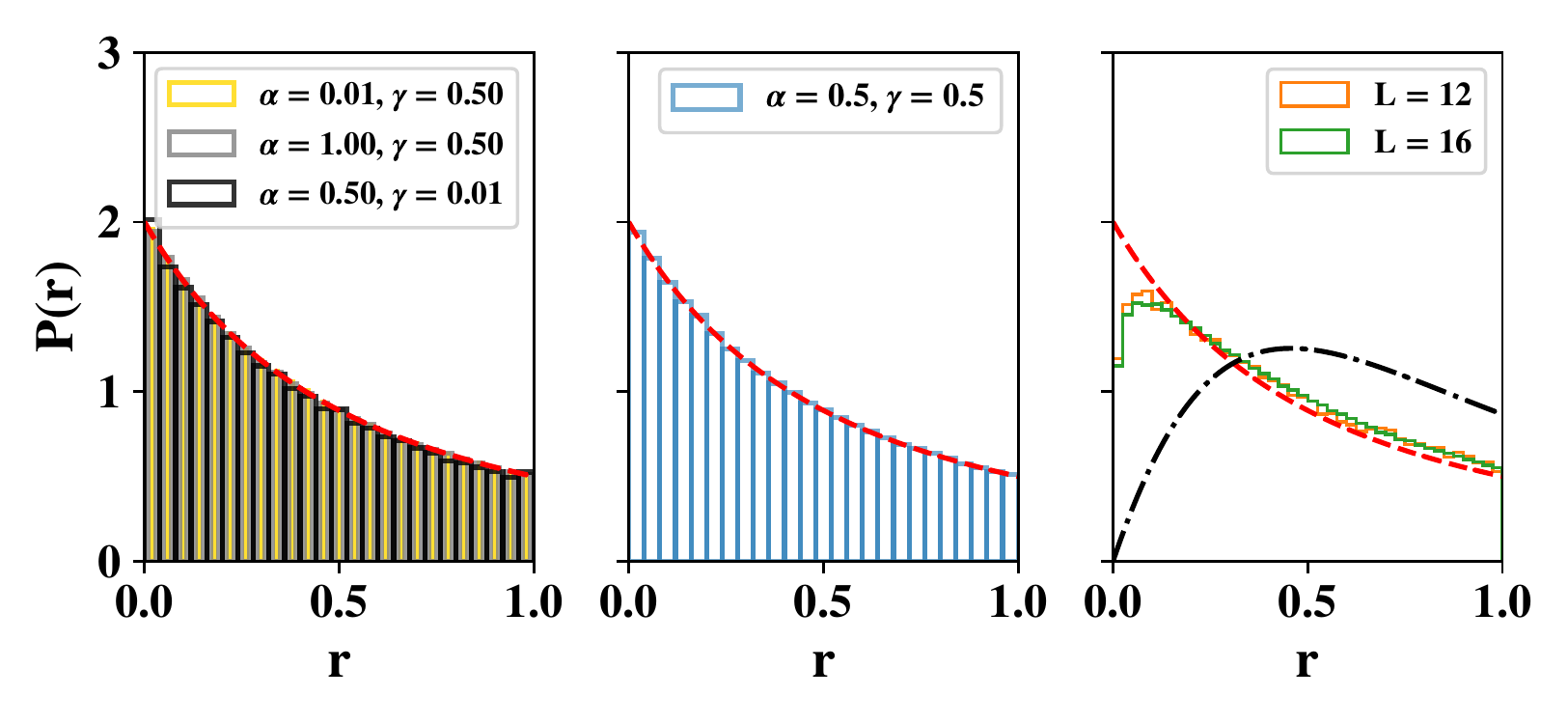}
\caption{ \textbf{Level-spacing ratio distributions.} Left: In the vicinity of the integrable lines, $P(r)$ agrees with that expected for a Poisson spectrum (red line). Center: Distributions remain indistinguishable from the Poisson one (red line) at moderate $\alpha$ and $\gamma$ when the system is no longer expected to be integrable. Right: At large disorder strength, we see level repulsion and a weak trend toward the Wigner-Dyson distribution (black line) with increasing $L$. Parameters: $\sum_{j} S_j^z=-1$, $N_s = 500$, $\omega_0=\alpha$, $L=16$ (left, center), $\gamma=10.0$ (right), $\alpha=0.5$ (right). Energies are sampled in the middle two quartiles of the spectrum. \label{fig:LStats}}
\vspace{-\baselineskip}
\end{figure}

Fig.~\ref{fig:LStats} shows the distribution $P(r)$ of the ratio $r$ of consecutive energy level spacings in a sector with fixed polarization. The ratio $r_n$ for the trio of energy levels with energies $E_{n\pm 1}, E_n$ is defined as $r_n = \min(s_n,s_{n-1})/\max(s_n,s_{n-1})$, where $s_n = E_{n+1} - E_{n}$ and the energy levels are ordered $E_1 < E_2 < E_3 < \ldots $~\cite{oganesyan2007localization, atas2013distribution}. The left panel shows that $P(r)$ agrees with that expected for a Poisson spectrum near/on the integrable lines points with $\alpha \approx 0$, $\alpha = 1$, and $\gamma \approx 0$. The center panel shows that the Poisson behavior persists in the presence of moderate anisotropy ($\alpha=0.5$) and disorder ($\gamma = 0.5$) at the largest size we access numerically. Only when the disorder strength is much larger than one ($\gamma=10.0$) do we see some level repulsion with a weak trend towards Wigner-Dyson statistics with increasing $L$ (right panel). Fig.~\ref{fig:LStats} therefore shows no tendency of the model to become chaotic with increasing $L$ at moderate values of $\alpha$ and $\gamma$. 

Recently, Ref.~\cite{pandey2020adiabatic} proposed the norm of the AGP as a highly sensitive probe for chaos. Related measures were also proposed earlier in the context of many-body localization, see e.g. Refs.~\cite{Serbyn_2015, crowley2019avalanche}. Chaos manifests in the exponential scaling of the Frobenius norm of the AGP with system size, which can be interpreted as an exponential sensitivity of the eigenstates to perturbations of the Hamiltonian. In contrast, integrable perturbations show polynomial scaling~\cite{pandey2020adiabatic}.

For any Hamiltonian $H(\alpha)$, the AGP operator can be represented as \cite{claeys2019floquet,pandey2020adiabatic}:
\begin{equation}
\mathcal{A}_{\alpha} = \lim_{\mu\to0^+} \int_{0}^{\infty} \,dt\, e^{-\mu t}\,\left( e^{-i H\,t}\,\partial_{\alpha} H\, e^{+i H \,t} - \mathcal{M}_{\alpha}\right),
\label{eq:AGP_integral}
\end{equation}
where $\mathcal{M}_{\alpha} \equiv \sum_n |n\rangle \langle n| \partial_{\alpha} H |n\rangle \langle n| $.
In the energy eigenbasis of $H(\alpha)$, the off-diagonal matrix elements of $\mathcal{A}_{\alpha}$ read:
\begin{equation}
\langle m |\mathcal{A}_{\alpha} |n\rangle = i \langle m | \partial_{\alpha}n \rangle = \lim_{\mu\to0^+} \frac{\langle m |\partial_{\alpha}H |n\rangle}{\mu+i\,\Omega_{mn}}.
\end{equation}
where $\Omega_{mn} = E_m - E_n$. Note that the diagonal matrix elements $\langle m |\mathcal{A}_{\alpha} |m\rangle = 0$, which is a gauge choice~\cite{Kolodrubetz}. The (scaled) Frobenius norm is then given by:
\begin{equation}\label{eq:AGPnorm}
\|\mathcal{A}_{\alpha} \|^2 = \frac{1}{2^{L}} \lim_{\mu\to0^+}  \sum_{m\neq n}   \frac{ |\langle m |\partial_{\alpha}H |n\rangle|^2}{\mu^2 + \Omega_{mn}^2}.
\end{equation}

In chaotic systems, $\|\mathcal{A}_{\alpha} \|^2$ fluctuates wildly with $L$ when we take the limit $\mu\to0^{+}$ because the terms with the smallest energy differences $\Omega_{mn}$ dominate the sum in the norm. This is a standard manifestation of the problem of small denominators~\cite{prigogine1991integrability}. Instead of taking the limit $\mu\to0^+$, it is convenient to set 
$\mu > 0$ as a regulator. This regulator provides a two-fold advantage: (i) it suppresses the wild fluctuations of the norm with system size, and (ii) it allows us to retain the exponential sensitivity of the AGP norm to small perturbations if we pick $\mu = L\, 2^{-L}/c$, where $c$ is a system-size-independent constant. The regulator $\mu$ is thus parametrically larger than the level spacing, while maintaining a small deviation from the exact AGP~\cite{pandey2020adiabatic}. Physically, $\mu^{-1}$ plays the role of a cutoff time for operator growth in Eq.~\eqref{eq:AGP_integral}. By picking this time to be exponentially large in $L$, we probe the sensitivity of eigenstates to infinitesimal perturbations.

For systems satisfying the ETH~\cite{d2016quantum} the states at energy density corresponding to infinite temperature satisfy
\begin{equation}
    |\langle m |\partial_{\alpha}H\, |n\rangle|^2=\frac{ R_{mn}^2}{2^L} |f(\Omega_{mn})|^2 ,
    \label{eq:offdiagonal_ETH}
\end{equation}
where $R_{mn}$ is a random variable with zero mean and unit variance, and $f(\Omega)$ is a smooth function that is proportional to the Fourier transform of the correlation function $\Tr(\partial_{\alpha}H(t)\partial_{\alpha}H(0)+\partial_{\alpha}H(0)\partial_{\alpha}H(t))$ at the frequency $\Omega$. In general, the function $f$ also depends on the average energy $(E_m+E_n)/2$. However, as the summation in Eq.~(\ref{eq:AGPnorm}) is dominated by the eigenstates corresponding to infinite temperature, we suppress this additional dependence. Typically, $|f(\Omega)|^2$ increases as $\Omega$ decreases until $1/\Omega$ becomes comparable to the slowest relaxation time scale in the system (such as the Thouless time) and saturates for smaller values of $\Omega$. 

Interestingly, it was recently observed that the function $|f(\Omega)|^2$ can be defined and remains smooth even in generic integrable systems~\cite{leblond2019entanglement}. Then $|f(\Omega)|^2$ vanishes as $\Omega\to0$ for deformations of the Hamiltonian along integrable directions~\cite{pandey2020adiabatic, Dymarsky_2019, brenes2020eigenstate,brenes2020ballistic}.

Combining Eq.~\eqref{eq:AGPnorm} at finite $\mu$ and Eq.~\eqref{eq:offdiagonal_ETH} at exponentially small scales $\Omega = \mu \sim L\,2^{-L}$, we get the following estimate for the AGP norm:
\begin{equation}
    \|\mathcal A_{\alpha}\|^2\sim \frac{|f(\mu)|^2}{\mu}.
    \label{eq:AGP_norm_ETH}
\end{equation}
Thus if our system were to satisfy ETH,
\begin{align}
    \|\mathcal A_\alpha\|^2 \sim 2^L \quad \textrm{(ETH)}, \label{Eq:AGPNormETH}
\end{align}
up to polynomial corrections. In contrast, we would expect only a polynomial scaling of $\|\mathcal A_\alpha\|^2$ with the system size in  an (interacting) integrable system for deformations of the Hamiltonian along an integrable direction~\cite{pandey2020adiabatic}.

\begin{figure}[tp]
\includegraphics[width=\columnwidth]{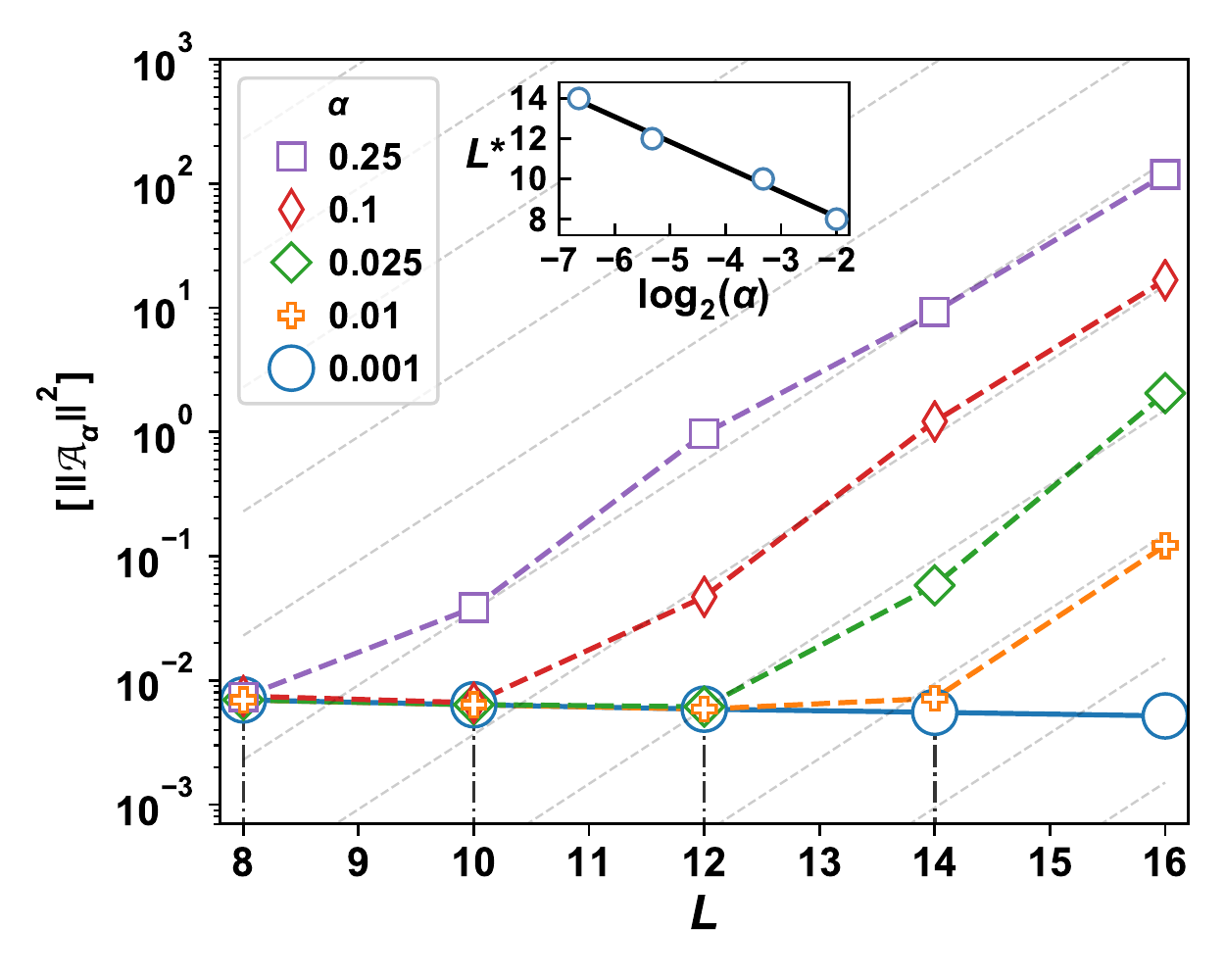}
\caption{ \textbf{The exponential divergence of the adiabatic gauge potential norm shows signatures of chaos.} Plot shows the disorder-averaged norm $[\,\|\mathcal{A}_{\alpha}\|^2]$ as a function of system size $L$. Dotted lines show the scaling behavior in the chaotic non-ergodic regime $[\,\|\mathcal{A}_{\alpha}\|^2] \sim 2^{2L}$. Vertical dashed-dot lines mark the onset of exponential growth at $L^{*}(\alpha)$. Inset: $L^{*}(\alpha)$ vs. $\log_2 (\alpha)$, and a regression line whose slope is numerically found to be $-\nu \approx -1.25$. Parameters: $\gamma = 0.5$, $N_s=200$, $\omega_0 = \alpha$, $\sum_{j} S_j^z=-1$, $\mu/L = 2^{-L}/c$, with $c\approx15$. \label{fig:AGPnorm}}
\vspace{-0.2\baselineskip}
\end{figure}

\begin{figure}[tp]
\includegraphics[width=\columnwidth]{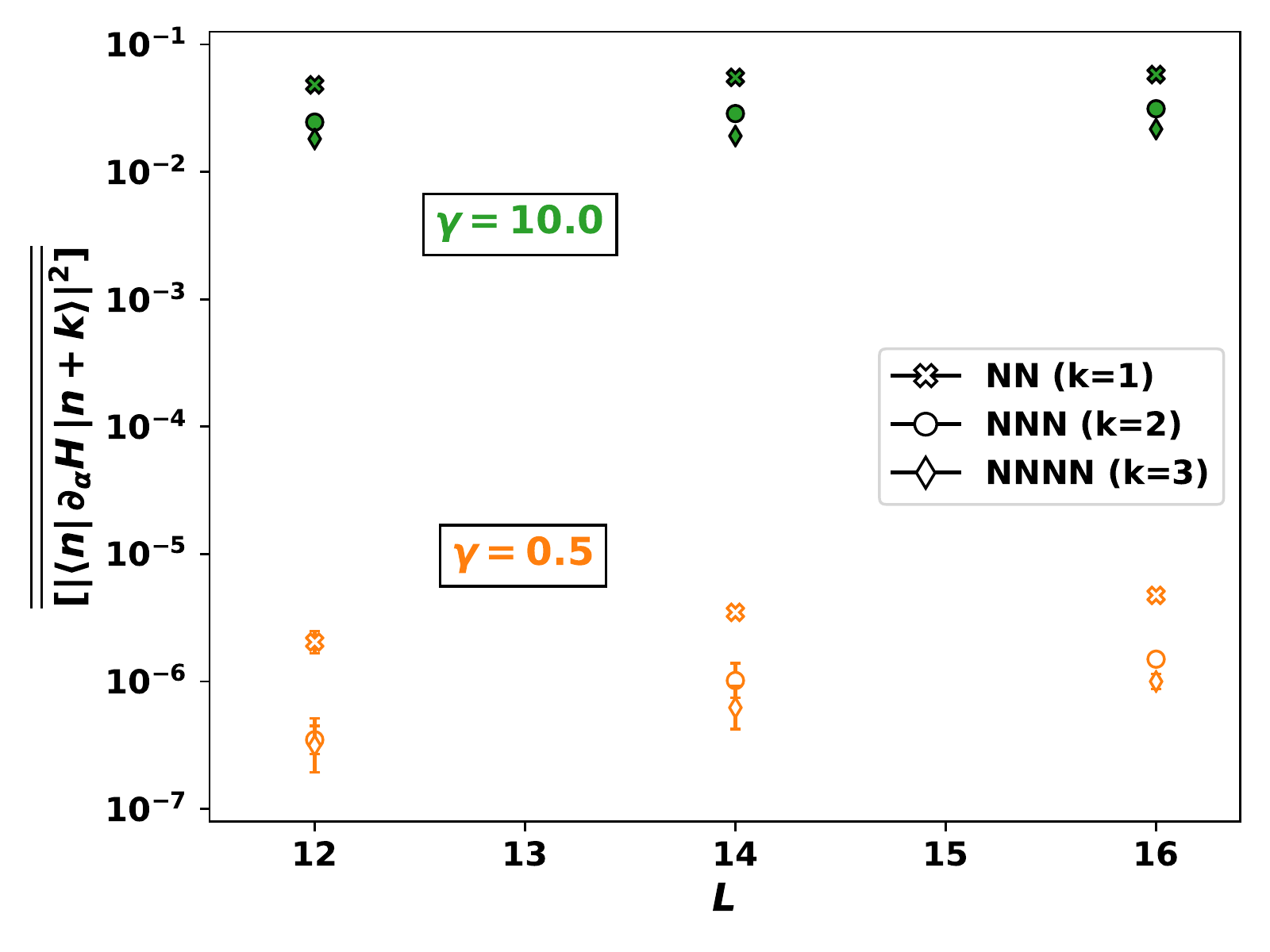}
\caption{\textbf{Nearest neighbor matrix elements of $\partial_{\alpha} H$ do not decay with system size $L$.} Plot shows nearest neighbor (NN), next nearest neighbor (NNN), and next next nearest neigbor (NNNN) matrix elements (squared) averaged over eigenstates $|n\rangle$ and disorder. Green filled markers ($\gamma=10.0$) and orange open markers ($\gamma=0.5$) show no decay with system size in the chaotic non-ergodic regime.  Parameters: $\alpha=0.5$, $\omega_0 = \alpha$, $\sum_j S_j^z=-1$, $N_s=1000$. Eigenstates $|n\rangle$ are sampled in the energy window $E_n \in [-0.5\,E_{\mathrm{max}},0.5 \,E_{\mathrm{max}}]$, where $E_\mathrm{max}$ is the maximum energy of $H$ in the given magnetization sector.\label{fig:NNMatelems}}
\vspace{-0.0\baselineskip}
\end{figure}

Fig.~\ref{fig:AGPnorm} shows the exponential divergence of the disorder-averaged norm $[\|\mathcal{A}_{\alpha}\|^2]$ of the AGP corresponding to perturbations of the anisotropy parameter $\alpha$ at moderate disorder strength $\gamma=0.5$. At $\alpha=0$, $[\|\mathcal{A}_{\alpha} \|^2]$ scales polynomially with system size $L$. Away from the integrable point, the scaling of $[\|\mathcal{A}_{\alpha} \|^2] $ is polynomial until a critical length $L^*(\alpha)$, which marks the onset of exponential growth and thus chaos (see vertical dash-dotted lines).  The critical length increases with decreasing $\alpha$ as:
\begin{align}
    L^*(\alpha) \sim -\nu \log_2 \alpha
    \label{Eq:Lstaralpha},
\end{align}
such that $L^{*}$ becomes infinite in the integrable limit $\alpha \to 0$. The power $\nu \approx 1.25$ is found using linear regression (see inset).  

For $L$ larger than the critical length $L^*(\alpha)$, the norm of the AGP in Fig.~\ref{fig:AGPnorm} grows exponentially at twice the rate predicted by the ETH:
\begin{align}
    \|\mathcal{A}_{\alpha} \|^2 \sim 2^{2(L-L^*(\alpha))} \quad \textrm{(CNE)}\label{eq:weakchaosscaling}.
\end{align}
From Eq.~\eqref{eq:AGP_norm_ETH}, we obtain $|f(\mu)|^2\sim 1/\mu\sim 2^L$. It follows from Eq.~(\ref{eq:offdiagonal_ETH}) that the off-diagonal matrix elements $\langle m |\partial_{\alpha} H |n \rangle$ are not exponentially suppressed with system size in the narrow energy interval $\Omega \sim \mu\sim 2^{-L}$, in contrast with the ETH prediction. The absence of an exponential suppression in the off-diagonal matrix elements is shown in Figure~\ref{fig:NNMatelems}. The figure shows nearest neighbor (in energy) matrix elements $|\langle n |\partial_{\alpha} H |n + k \rangle|
^2$ for $k\in\{1,2,3\}$, averaged over disorder samples and eigenstates $|n\rangle$; we denote this averaging by the double-overline $\overline{\overline{[\cdots]}}$. These matrix elements differ by several orders of magnitude between low ($\gamma=0.5$) and high ($\gamma=10.0$) disorder. However, no exponential decay with $L$ is observed at either disorder strength at numerically accessible systems sizes. This absence of exponential suppression can only persist when polynomially many nearby eigenstates mix. In contrast, ETH would require that a given eigenstate mix equally with exponentially many nearby eigenstates upon perturbing the system.

\begin{figure}[bp]
\includegraphics[width=\columnwidth]{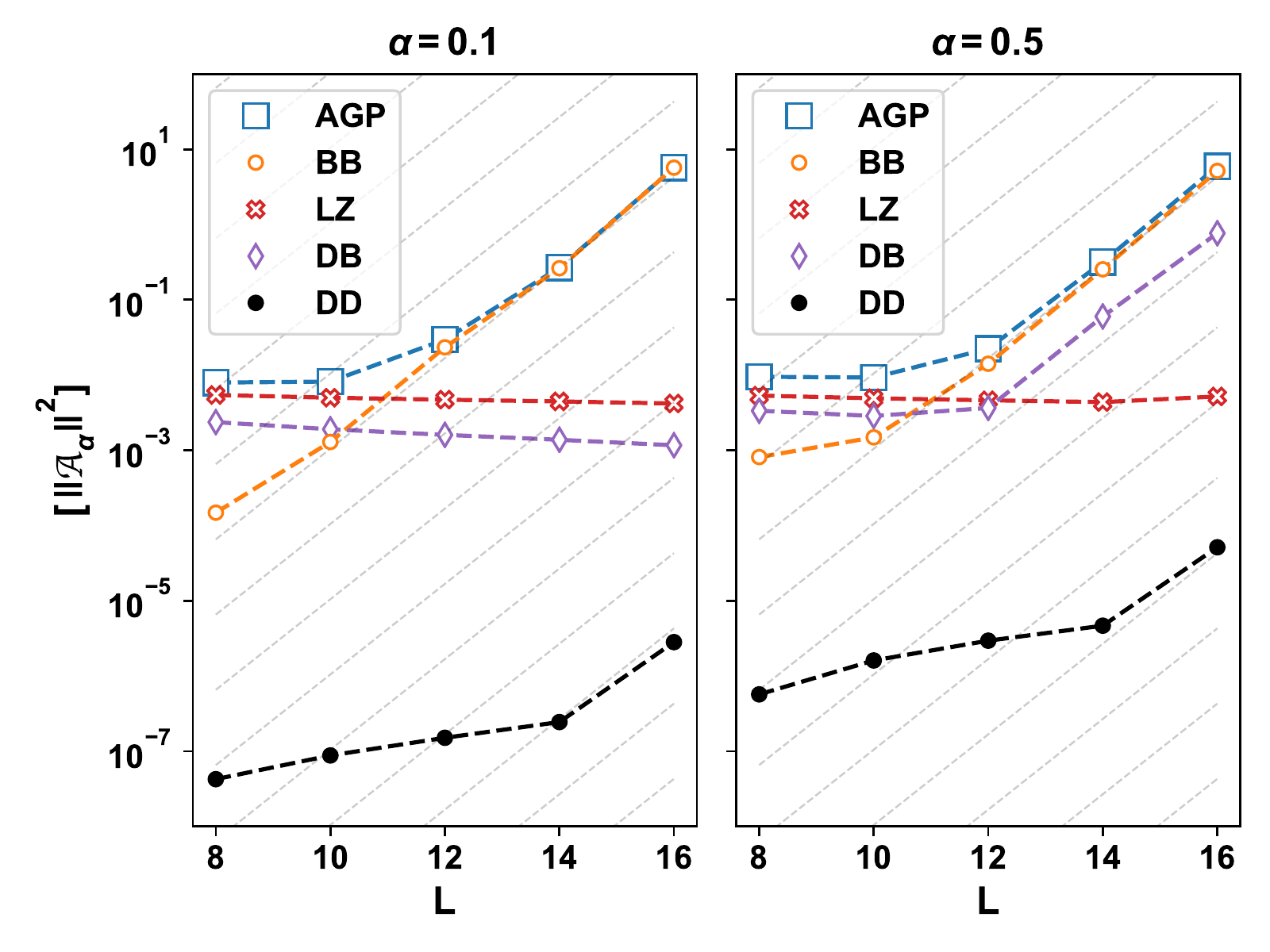}
\caption{\textbf{ Decomposition of the adiabatic gauge norm into classes based on the nature of the eigenstates.} The disorder-averaged AGP norm (AGP) is composed of contributions from matrix elements between dark states (DD), bright states in the same band (BB), bright states in different bands (LZ), and dark and bright states (DB). Left: At $\alpha=0.1$, BB and DD show onset of non-ergodic chaos. The AGP norm is dominated by BB, whereas DD is negligible. As DB is relatively constant with $L$, dark states do not progressively hybridize with bright states at accessible sizes. Right: At $\alpha=0.5$, the AGP norm is still dominated by BB. However, dark states hybridize strongly with nearby bright states due to the exponential divergence of DB contribution. Parameters: $\gamma=0.5$, $\omega_0 = \alpha$, $\sum_j S_j^z=-1$, $N_s=1000$, cutoff $\mu/L= 2^{-L}/c$ with $c=6$ (left) and $c=0.7$ (right). The values of $c$ in the two panels are chosen such that $L^{*} \approx 10$. \label{fig:AGPnorm_Breakdown}}
\vspace{-0.3\baselineskip}
\end{figure}

The behavior of the chaotic non-ergodic regime is manifest separately within the dark and bright manifolds, and jointly in the interactions between these manifolds. We plot the various contributions to the disorder-averaged AGP norm (AGP) in Eq.~\eqref{eq:AGPnorm} from dark and bright classes of eigenstates in Fig.~\ref{fig:AGPnorm_Breakdown}. The DD (BB) contribution comes from terms involving matrix elements between dark eigenstates (bright eigenstates in the same band\footnote{The bright states in the XX model come in Landau-Zener (LZ) pairs that can be continuously followed as a function of $\omega_0$ in each magnetization sector. These bright states form two bands, consisting of the positive and negative energy states of each LZ pair, respectively. See the Supplemental Information.}). The matrix elements between the dark and bright states contribute the DB piece, while the matrix elements between the two bright state bands contribute the Landau-Zenner (LZ) piece. At $\alpha=0.1$ and $\alpha=0.5$, the sum in the AGP norm is dominated by the intra-band bright-bright (BB) matrix elements. Consequently, the off-diagonal matrix elements between neighboring bright states are not exponentially suppressed (see the discussion below Eq.~\eqref{eq:weakchaosscaling}). Dark-dark contributions also exponentially increase with $L$; however their total value is many orders of magnitude smaller than the BB contribution at these sizes. The DB contributions show a striking difference between the left and right panels of Fig.~\ref{fig:AGPnorm_Breakdown} at the accessible sizes. The DB contributions only grow exponentially with $L$ in the right panel with $\alpha=0.5$, reflecting the strong hybridization between neighboring dark and bright states in the spectrum on perturbing $\alpha$. For $\alpha=0.1$, on the other hand, the dark and bright state manifolds are separated in energy at the accessible sizes. This limits the hybridization between the two manifolds. However, we expect that the DB contribution diverges exponentially with $L$ at sufficiently large sizes at any $\alpha>0$. We remind the reader that perturbation theory in Sec.~\ref{sec:persistentdarkstates} worked exceedingly well to characterize qubit observables in dark states, even in parameter regimes where the DB contribution exponentially increases with $L$. This suggests that the strong DB mixing should primarily affect bath observables in persistent dark states at these sizes. In the next section, we discuss the potential implications of this behavior in the thermodynamic limit.

In sum, the exponential divergence of $[\|\mathcal{A}_{\alpha} \|^2 ]$ provides strong evidence that arbitrarily small in $\alpha$ perturbations are integrability-breaking, with a growth rate that is twice that predicted by the ETH at numerically accessible system sizes. In this chaotic non-ergodic regime, eigenstates in exponentially small shells of order $\mu$ hybridize, with interaction matrix elements that show no exponential suppression with system size. 

\section{Possible fates of the CNE regime in the thermodynamic limit}
\label{sec:relaxation}
We have provided evidence that the family of central spin models in Eq.~\eqref{eq:H} is chaotic, but non-ergodic at moderate disorder strengths and numerically accessible system sizes. Eigenstate chaos manifests in the exponential scaling of the AGP norm with $L$, while the non-ergodicity is manifest in both the non-thermal value of the central spin polarization in persistent dark states and the non-ETH scaling of the AGP norm. 

Ref.~\cite{pandey2020adiabatic} argued that the non-ETH scaling of the AGP norm in Eq.~\eqref{eq:weakchaosscaling} indicates slow relaxation with exponentially long in $L$ relaxation times. We repeat the argument for completeness. From Eq.~\eqref{eq:AGP_norm_ETH} we find:
\begin{align}
 |f(\mu)|^2 \propto \mu \|\mathcal A_\alpha\|^2.
\end{align}
As $\Omega \to 0$, $|f(\Omega)|^2$ is proportional to the relaxation time $\tau_r$ of the operator $\partial_\alpha H=\sum_i g_i S_0^z S_i^z$. Using that $\mu \propto L \,2^{-L}$ and Eq.~\eqref{Eq:Lstaralpha}, we find that for $L>L^\ast(\alpha)$:
\begin{equation}
    \tau_r \sim C |\alpha|^{2\nu} 2^L, \label{Eq:taurweak}
\end{equation}
where $2\nu\approx 2.5$ and the constant $C$ can have a weak (power law) dependence on $L$. As the DD, DB and BB components of the AGP norm exhibit non-ETH scaling in the right panel of Fig.~\ref{fig:AGPnorm_Breakdown}, we expect that Eq.~\eqref{Eq:taurweak} characterizes certain relaxation processes in both the dark and bright sectors.

In the dark eigensector, the exponentially long relaxation times largely arise from the dark-bright mixing (cf. Fig.~\ref{fig:AGPnorm_Breakdown}) and coexist with a robust non-thermal value of the central spin magnetization. This suggests the following `cartoon' for the decomposition of the persistent dark states in the eigenbasis of the XX integrable model,
\begin{equation}
    |\mathcal D(\alpha) \rangle=\sqrt{Z}\, |\uparrow\,\rangle \otimes |\mathcal D^+\rangle+\sqrt{1-Z} \,|\tilde{\mathcal{B}}\rangle,
\label{eq:wave_function_Z}
\end{equation}
with a non-zero central spin residue $Z \in (0,1]$. Above, we use the eigenbasis of the XX model at resonance ($\omega_0 =0$) in a magnetization sector with fixed positive value, $|\mathcal{D}^+\rangle$ is a dark states satisfying Eq.~\eqref{eq:DarkStates}, and $|\tilde{\mathcal{B}}\rangle$ is a normalized superposition of bright states with the property $\langle\mathcal{B}_i|S_0^z|\mathcal{B}_j\rangle = 0$~\footnote{In the XX model, $S_0^z$ only connects pairs of bright states with equal and opposite energy~\cite{villazon2020integrability}. As the bright states hybridize in (exponentially small in $L$) energy shells due to the $\alpha$-perturbation, it is plausible that $\ket{\tilde{B}}$ statistically does not involve these pairs.}. From this property, we obtain:
\begin{align}
\langle \mathcal D(\alpha)|S_0^z  |\mathcal D(\alpha) \rangle \approx \frac{Z}{2}. 
\end{align}
In contrast, the central spin polarization equals $\sum_j S_j^z/L$ in an infinite temperature eigenstate of the thermalizing system, which vanishes as $L^{-1}$ in a fixed magnetization sector.

\begin{figure}[tb]
\centering
\includegraphics[width=\columnwidth]{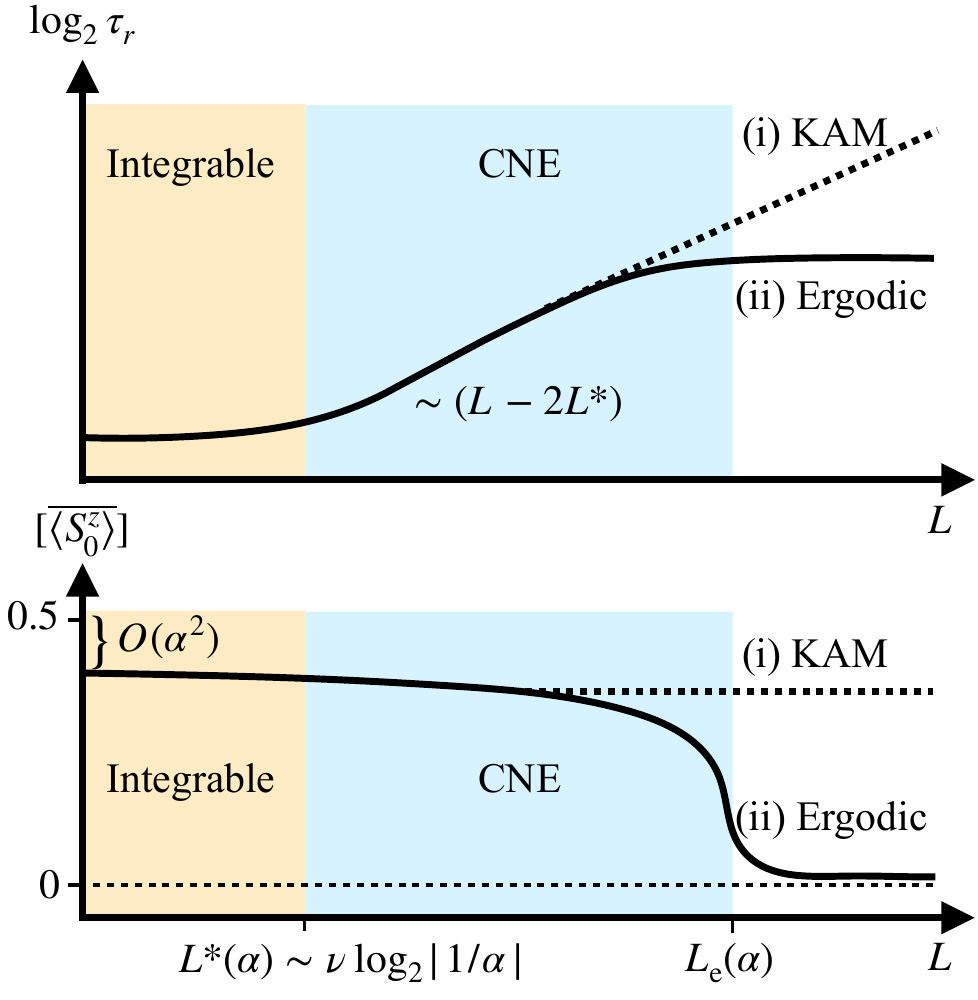}
\caption{ \textbf{ Schematic of possible scenarios in the thermodynamic limit.} Upper panel shows (log) relaxation time $\tau_r$ vs $L$, while lower panel shows disordered averaged central spin magnetization in the dark manifold $[\overline{\langle\ {S_0^z}\rangle}]$ vs $L$. The integrable regime (left) is controlled by the integrable lines. In the CNE regime (center), the relaxation time in the bath is exponentially long in $L$, despite the persistence of qubit polarization in dark states. As $L\to\infty$ (right), two scenarios are shown: (i) KAM (dotted curves), where the CNE regime persists at all system sizes and (ii) Ergodic (solid curve), where the CNE regime crosses over to the normal ETH ergodic behavior.  \label{fig:Fates}}
\vspace{-\baselineskip}
\end{figure}

Fig.~\ref{fig:pt} indicates that at intermediate disorder $\gamma=0.5$, the residue $Z$ is well-captured by perturbation theory with no noticeable system size dependence. For stronger disorder, we however find that $Z$ slightly decreases with $L$ (this can be inferred from the bottom panel of Fig.~\ref{fig:long-time}). This suggests that the (exponentially) long relaxation times are associated with slow dynamics of the bath spins adjusting to long time (likely non-thermal) configurations. Likewise, the exponential increase of the AGP norm in the dark sector is mostly due to their mixing with bright states (cf. Fig.~\ref{fig:AGPnorm_Breakdown}), in turn implying that the $|\tilde{\mathcal B}\rangle$ part of the wave function in Eq.~\eqref{eq:wave_function_Z} is chaotic, i.e. exponentially sensitive to infinitesimal perturbations.

From the presented data, it is not possible to predict what happens as the system size $L$ increases beyond $L\approx 16$. We propose two possible distinct possibilities:
\begin{enumerate}
\item KAM-type: the residue $Z$ remains finite, the bath remains non-ergodic, and $||\mathcal A_\alpha||^2\propto \exp[2\log(2) L]$ persists as $L \to \infty$.
 \item Ergodic: the residue $Z$ ultimately vanishes as $L\to \infty$ and the AGP norm crossovers to the ETH scaling, $||\mathcal A_\alpha||^2\propto \exp[\log(2) L]$.
 \end{enumerate}
One can also imagine other more exotic scenarios, where, for example, the residue $Z$ remains finite at $L\to\infty$ while the bath becomes ergodic, or conversely, $Z\to 0$ but the whole system remains non-ergodic. We do not discuss these further as we see no indications that they could be realized. 

Fig.~\ref{fig:Fates} schematically depicts the possible scenarios of (i) KAM and (ii) Ergodic behavior. The figure shows the system size dependence of the (log) relaxation time $\tau_r$ (upper panel) and the disorder averaged central spin magnetization in the dark manifold (lower panel). At system sizes smaller than the critical size $L^*(\alpha)$, we have a region where the dynamics of the system are dominated by the integrable lines and the system quickly relaxes to a non-thermal steady state, with at most polynomial dependence of $\tau_r$ on $L$. In the chaotic non-ergodic (CNE) regime, eigenstate mixing gives rise to an exponentially increasing relaxation time for the bath (cf. Eq.
~\eqref{Eq:taurweak}), while persistent dark states maintain a non-thermal qubit polarization. As we approach the thermodynamic limit $L\to\infty$, scenario (i) would result in a continuation of the CNE regime (see dotted curves), while (ii) would show a second critical size $L_e(\alpha)$ marking the onset of ergodic dynamics. For $L \gg L_e(\alpha)$, $\tau_r$ saturates and the system always reaches local thermal equilibrium under its own isolated dynamics.

Both possibilities outlined above are very interesting and have nontrivial implications. If the KAM scenario (i) is realized, then there is a true non-ergodic phase in the thermodynamic limit. In this case, both the dark and the bright sectors behave non-ergodically at all system sizes, violating the ETH. They are characterized by bath-spin relaxation times that are exponentially long in $L$. These violations will necessarily lead to a breakdown of various thermodynamic relations such as the fluctuation-dissipation theorem, which heavily rely on ETH~\cite{d2016quantum}. If the Ergodic scenario (ii) is realized, then the system will eventually relax in a finite time to a thermal steady state with a thermal value of the central-spin magnetization. Even if this scenario is realized, according to our numerical results, this can only happen at extremely large system sizes (cf. the lower panel in Fig.~\ref{fig:long-time}). As the relaxation time $\tau_r$ scales exponentially with $L$, it could be astronomically large at $L \approx L_e$ before saturation. This suggests that the dark states, while not exact eigenstates in scenario (ii), will be extremely stable and long lived.

Our numerical results do not predict which of the two possibilities is realized. Based on the available data the KAM scenario (i) seems to be the most likely, at least for moderate $\alpha$ and $\gamma$, as there are no visible deviations from $Z(L)=\textrm{const.}$ (Fig.~\ref{fig:pt}) and $||\mathcal A_\alpha||^2\propto \exp[2\log(2) L]$ (Fig.~\ref{fig:AGPnorm}).  The absence of deviations from these scalings is especially remarkable since there are no small parameters in the system, so there is no obvious estimate for the length scale $L_e$. Nevertheless, a more careful analysis is needed to reach a definite conclusion. 

\section{Discussion}
\label{sec:discussion}
Physically, dark states can be realized in several qubit systems with mesoscopic environments. For example, in diamond systems, a nitrogen vacancy (NV) center serves as a qubit and the electronic spins on the surface act as a bath.  In a suitable rotating frame, the Hamiltonian is well approximated by the XX central spin model. Furthermore, as the qubit-bath interactions are dipolar, they decay sufficiently rapidly as the distance between the NV and a surface spin grows that only a handful of surface spins can be experimentally accessed~\cite{hall2014analytic,rios2010quantum}. Our results imply that such small NV-surface spin systems exhibit dark states that are robust to the presence of moderate anisotropy and disorder. 

One potential avenue for applications involves quantum information processing in the manifold of persistent dark states. To initialize the system in the persistent dark state manifold, one can implement dynamic nuclear polarization (DNP) to repeatedly polarize the central spin and transfer its polarization to the bath. DNP works by harnessing the flip-flop interactions ($S_0^{+}S_j ^{-}+S_0^{-}S_j ^{+}$) already present in the Hamiltonian $H$. This transfer can be achieved with several methods, e.g. tuning external fields to resonant Hartmann-Hahn conditions where flip-flop interactions dominate~\cite{hartmann1962nuclear,rovnyak2008tutorial}. By Eq.~\eqref{eq:DarkStates}, dark states are unaffected by flip-flop interactions, and therefore only bright state populations will be continually transferred to dark states and other bright states. Repeating the process, the system tends to a statistical mixture of persistent dark states~\cite{gullans2013preparation,christ2007quantum}. The low qubit-bath entanglement of the resulting manifold ensures robust qubit states with large decoherence times for high-fidelity quantum computing.

A closely related application of DNP is to fully polarize a mesoscopic bath. It has long been known that DNP protocols populate dark states in which the bath is only partially polarized, preventing complete bath polarization and severely limiting this goal~\cite{urbaszek2013nuclear}. Methods to overcome these limitations have been proposed~\cite{christ_nuclear_2009,imamoglu_optical_2003}. Our results extend these limitations to mesoscopic central spin systems with moderate anisotropy and disorder. A promising avenue for future research would be to characterize experimentally relevant integrability-breaking perturbations which destroy mesoscopic persistent dark states. 

Our results also extend the class of mesoscopic systems relevant for applications to quantum memory. 
Dark states have been proposed for the storage and retrieval of qubit states~\cite{taylor_controlling_2003,taylor2003long}. In one scheme, the qubit is initialized in an arbitrary state, which can be expressed as a superposition of bright and dark states. By controlling the external field which does not couple to dark states, the information about the qubit state can be completely transferred to the surrounding bath state and retrieved at a later time~\cite{taylor_controlling_2003}. The scheme immediately generalizes to persistent dark states with moderate anisotropy and disorder, opening avenues for quantum memory in new systems. 

To conclude, we investigated the robustness of dark states in a family of central spin models with anisotropic and inhomogenous qubit-bath interactions. 
The model is integrable along three lines in parameter space, two of which exhibit exact \exact dark eigenstates in which the central spin is unentangled with its environment. 
At moderate deviations away from these exact lines, we found persistent dark states whose central spin polarization and entanglement entropy are well-described by perturbation theory at numerically accessible system sizes.
We furthermore showed that the extensive set of conserved operators at the integrable lines morph into an extensive set of quasi-conserved operators away from the integrable lines.
In quench experiments, these quasi-conserved operators result in non-thermal correlations in a long-lived non-thermal state.
To address the possibility of chaotic behavior at larger system sizes than numerically accessible, we investigated the scaling behavior of the norm of the generator of adiabatic deformations of eigenstates with system size.
Although the scaling predicts the onset of chaos at any non-zero strength of the (integrability-breaking) perturbation, the ETH is not obeyed and the relaxation time of the system diverges exponentially with system size. While these effects may disappear in thermodynamically large systems, we see no evidence for this at the numerically-accessible system sizes.

\section*{Acknowledgements.} 
The authors thank D. Sels, C.R. Laumann and A. Sushkov for insightful discussions and collaborations on related topics, and M. Rigol for useful feedback on the manuscript. 
The authors acknowledge support from the Sloan Foundation through a Sloan Research Fellowship (A.C.), from the Belgian American Educational Foundation (BAEF) through the Francqui Foundation Fellowship (P.W.C.), from the Banco Santander Boston University-National University of Singapore grant (M.P.), and from the BU CMT Visitor Program (P.W.C.). Numerics were performed on the BU Shared Computing Cluster with the support of the BU Research Computing Services.
This work was supported by EPSRC Grant No. EP/P034616/1 (P.W.C.), NSF DMR-1813499 (T.V. and A.P.) and NSF DMR-1752759 (T.V. and A.C.), and AFOSR FA9550-16- 1-0334 (A.P.).

\bibliography{}

\end{document}


\onecolumngrid

\author{Tamiro Villazon}
\email{rtvs@bu.edu}
\affiliation{Department of Physics, Boston University, 590 Commonwealth Ave., Boston, MA 02215, USA}

\author{Pieter W. Claeys}
\affiliation{TCM Group, Cavendish Laboratory, University of Cambridge, Cambridge CB3 0HE, UK}

\author{Mohit Pandey}
\affiliation{Department of Physics, Boston University, 590 Commonwealth Ave., Boston, MA 02215, USA}

\author{Anatoli Polkovnikov}
\affiliation{Department of Physics, Boston University, 590 Commonwealth Ave., Boston, MA 02215, USA}

\author{Anushya Chandran}
\affiliation{Department of Physics, Boston University, 590 Commonwealth Ave., Boston, MA 02215, USA}

\title{%
  Persistent dark states in anisotropic central spin models: \\
  \large Supplementary Information}

\nopagebreak
\maketitle

\section{Energy Spectrum and Central Spin Entanglement Off Resonance}

The main text introduces the Hamiltonian with a local magnetic field $\omega_0$ on the central spin, and a global field $\omega$ on the bath:
\begin{equation}\label{eq:H}
H = \omega_0 \,S_0^z +   \omega \sum_{i=1}^{L-1} S_i^z + \sum_{i=1}^{L-1} \,g_i \left( S_0^x\, S_i^x + S_0^y\, S_i^y + \alpha\,S_0^z\, S_i^z \right),
\end{equation}
where $\alpha$ tunes the qubit-bath interaction anisotropy and $g_i$ describes the qubit-bath interaction strengths $g_i = g_0(1 + \gamma\,\delta_i)$. The total magnetization $\sum_{j=0}^{L-1} \,S_j^z$ commutes with $H$, giving rise to polarization sectors with definite total magnetization.

This model has a natural resonance point condition where exchange interactions between the central spin and the bath are strongly enhanced. At $\alpha = 0$, resonance occurs when $\omega_0 = \omega$. At finite $\alpha>0$ and in a fixed polarization sector, the last term in $H$ shifts the resonance point since $\alpha\,g_0\, S_0^z \sum_{j=1}^{L-1} S_j^z = \alpha\,g_0\, S_0^z ( \sum_{j=0}^{L-1} S_j^z - S_0^z) = \alpha\,g_0\, S_0^z  \sum_{j=0}^{L-1} S_j^z - h_0 $, where $h_0 = \alpha g_0 (S_0^z)^2 = \alpha g_0 /4$ is a constant for central spin $1/2$. Collecting the terms in the Hamiltonian coupled to the central spin $S_0^z$ yields the shifted resonance condition:
\begin{equation}
\tilde{\omega}_0 \equiv \omega_0 + \alpha\,g_0\, \sum_{j=0}^{L-1} \,S_j^z = \omega . \end{equation}
Without loss of generality, we set $\omega = 0$ throughout this work, such that resonance is given by $\tilde{\omega}_0 = 0 \implies \omega_0 = - \alpha\,g_0\, \sum_{j=0}^{L-1} \,S_j^z$.

\begin{figure}[bp]
\includegraphics[width=1.0\columnwidth]{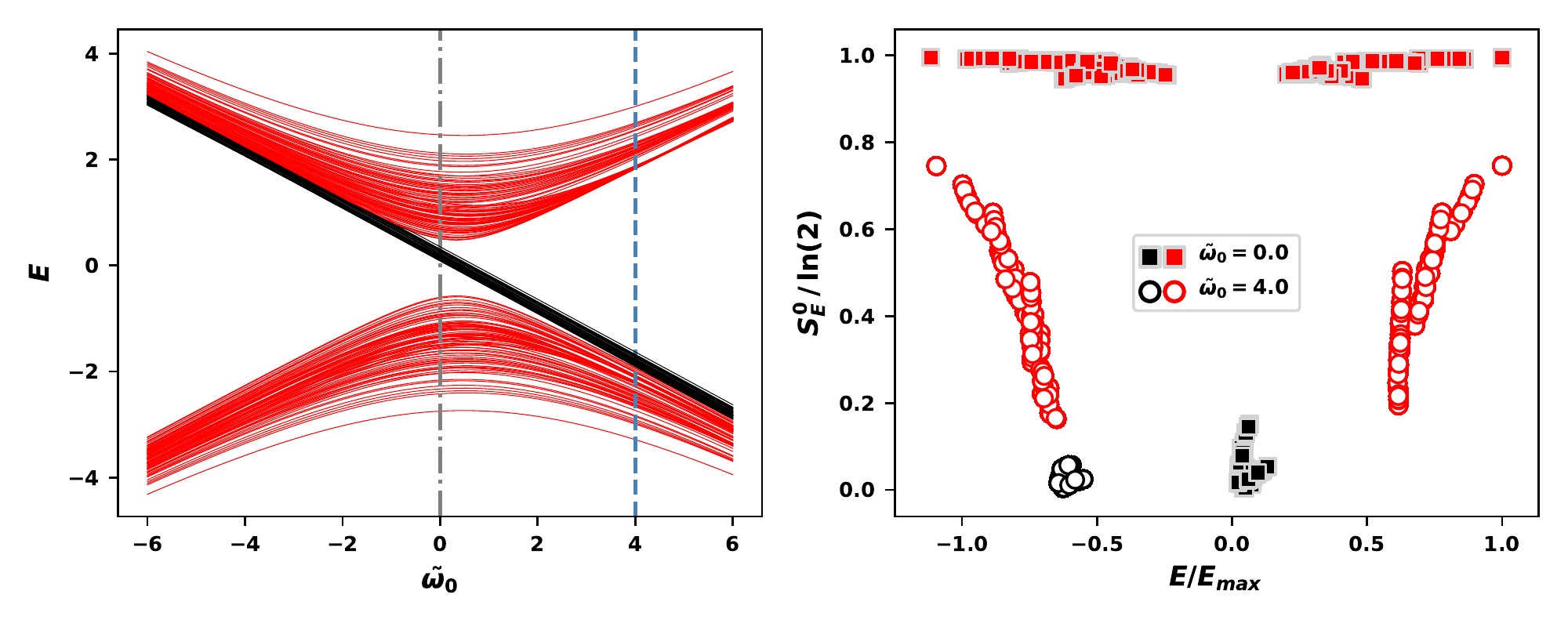}
\caption{ \textbf{Spectrum and central spin entanglement on and off resonance}. Left panel plots the energy $E$ as a function of the effective central field $\tilde{\omega}_0$, for all eigenstates of $H$. Right panel shows the central spin entanglement for all eigenvalues on resonance ($\tilde{\omega}_{0}=0$) and off resonance ($\tilde{\omega}_{0}=4$).
Vertical lines in the left panel denote the field values at resonance (gray dash-dotted line) and off resonance (blue dashed) used in the right panel. On resonance, dark and bright states can be easily distinguished by $E$ and $\mathcal{S}_E^0$, while off resonance these observables become comparable. Parameters: $L=11$, $N_s=1$ (typical sample), $\alpha=0.5$, $\gamma=0.5$, $\sum_j S_j^z = -0.5$. \label{fig:EEE}}
\vspace{-\baselineskip}
\end{figure}

The results shown in the main text focus on the physics of the system near resonance, where the difference between bright and dark states is most pronounced. This distinction is most clearly seen in the XX limit $(\alpha=0)$, where dark states are product states $\ket{\downarrow}_0 \otimes \ket{\mathcal{D}^-}$ or $\ket{\uparrow}_0 \otimes \ket{\mathcal{D}^+}$, whereas bright states have the form $c_1(\omega_0) \ket{\downarrow}_0 \otimes \ket{\mathcal{B}_{\downarrow}} + c_2(\omega_0) \ket{\uparrow}_0 \otimes \ket{\mathcal{B}_{\uparrow}}$, with nonzero $c_1$ and $c_2$ dependent on $\omega_0$. A thorough discussion of the spectrum in the XX limit is given in Ref.~\cite{villazon2020integrability}. Dark states are insensitive to changes in $\omega_0$. In contrast, bright states can be tuned to equal superpositions of the central spin up and down at resonance ($\omega_0=0$), or configurations where the central spin is mostly polarized along either direction (as $\omega_0 \to\infty$, $c_1\to0, c_2\to 1$ and as $\omega_0\to-\infty$, $c_1\to1, c_2\to 0$). Thus the central spin can be essentially decoupled from the bath in bright states with strong off-resonance fields. 

Figure~\ref{fig:EEE} shows the energy spectrum of $H$ (left panel) across a range of shifted central fields $\tilde{\omega}_0$, and the central spin entanglement entropy (right panel) for $\tilde{\omega}_0 = 0$ (squares) and $\tilde{\omega}_0 = 4$ (circles) -- see vertical dash-dotted and dashed lines in the left panel respectively. We have fixed total magnetization to $\sum_{j=0}^{L} S_j^z = -1 < 0$, such that dark states have $\langle S_0^z \rangle \approx -0.5$. In the spectrum, bright states come in pairs exhibiting level repulsion at resonance (see bands of red/light curves). Dark states show up as linear bands of near degenerate states (see black lines). 
Far from resonance, bright states attain nearly polarized central spins, and therefore lower central spin entanglement (as $\tilde{\omega}_0\to\pm\infty$, their entanglement approaches the entanglement of dark states). Thus the distinction between dark and bright states (as measured by observables such as $E, \langle S_0^z \rangle$, and $ S_E^0$) becomes progressively less sharp away from resonance, and must be characterized by alternative means (e.g. by their sensitivity to $\omega_0$).   

\section{Central Spin Projection: Breakdown of Perturbation Theory}

\begin{figure}[htp]
\begin{minipage}[t]{0.499\linewidth}
\includegraphics[width=1.0\linewidth]{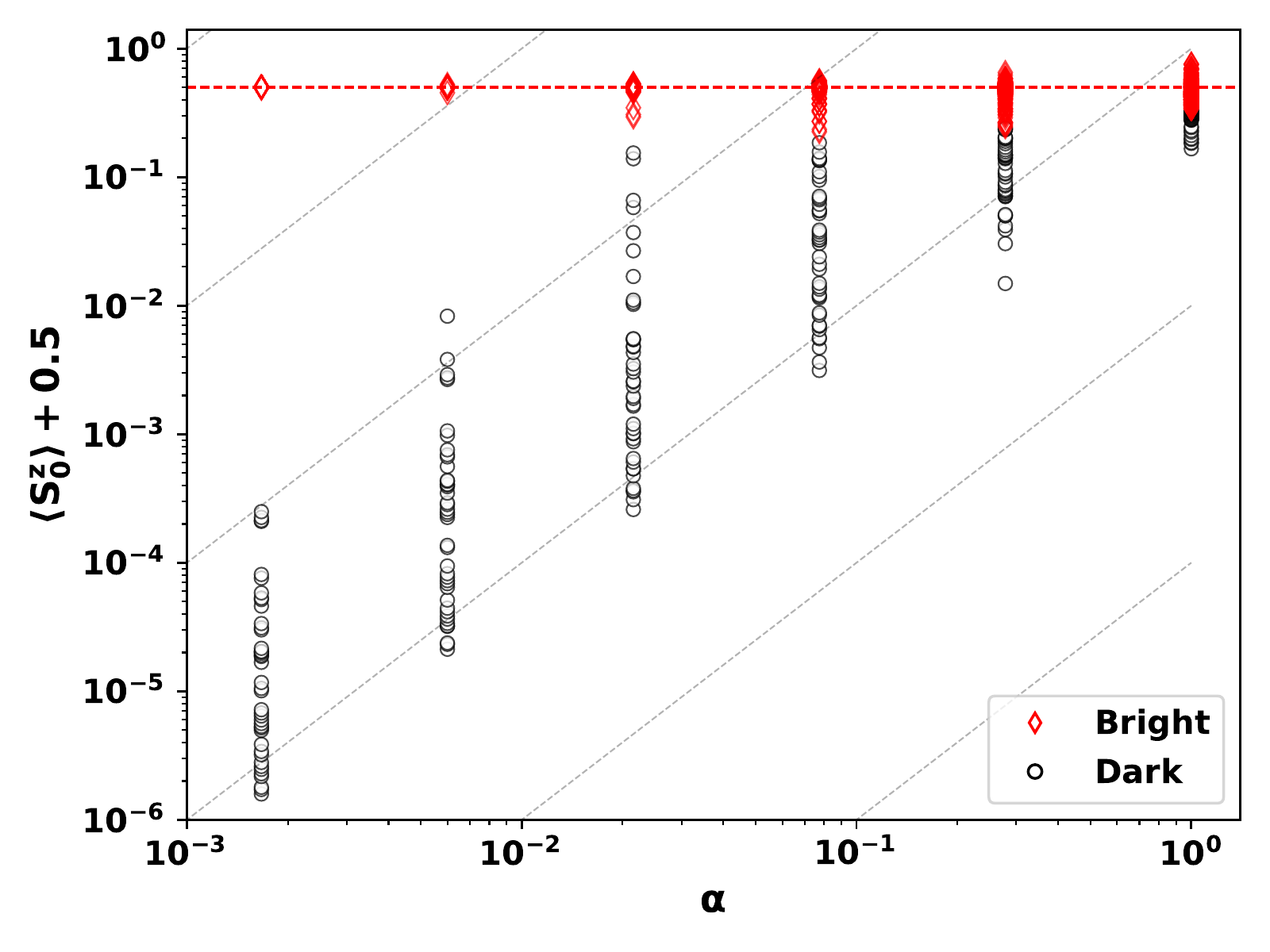}
\end{minipage}
\begin{minipage}[t]{0.499\linewidth}
\includegraphics[width=1.0\linewidth]{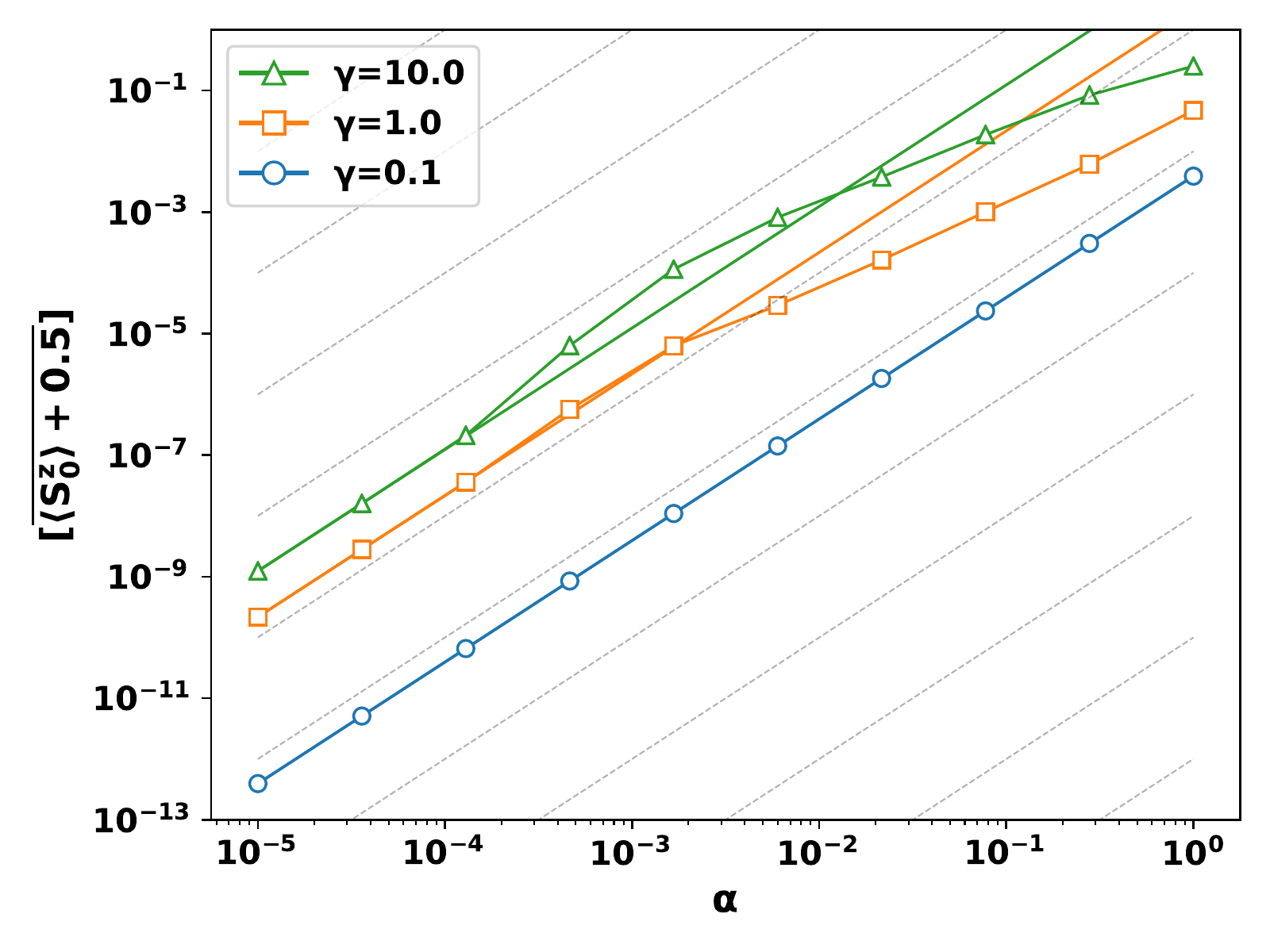}
\end{minipage}%
\caption{\textbf{Perturbation theory breaks down rapidly at large $\gamma$.} Left plot shows the expectation value of the central spin $z$-projection $S_0^z$ for a typical sample of disorder with strength $\gamma=10.0$. We see deviations from perturbation theory due to mixing between dark and bright states. The color coding used to separate dark and bright states is only nominal at sufficiently large $\alpha$, as the states can no longer be precisely separated into two distinct clusters. Right plot shows the expectation $\overline{[\langle S_0^z \rangle +0.5]}$ averaged over the $N_{D}$ eigenstates with smallest central spin projection, and $N_s=500$ disorder samples. The numerical data (markers) with $\gamma =1.0$ and $\gamma=10.0$ showcase this breakdown, as they deviate from their corresponding perturbation theory predictions (solid lines). Parameters $L=12$, $N_s=500$ (right), $\sum_j S_j^z = -1$, $\omega=\alpha$.
\label{fig:breakPT}}
\end{figure}

In the main text, we established how perturbation theory captures the behavior of observables such as the central spin expectation value $\langle \mathcal{D}(\alpha,\gamma)| S_0^z |\mathcal{D}(\alpha,\gamma) \rangle$, for a broad range of anisotropies $\alpha$ and small to moderate disorder $\gamma$. When $\gamma \gtrsim 1.0$, perturbation theory breaks down more rapidly as we tune $\alpha$ away from the $\alpha =0$ integrable line. This is shown in Fig.~\ref{fig:breakPT}. 

\section{Locality of the Adiabatic Gauge Potential}

\begin{figure}[htp]
\includegraphics[width=0.55\columnwidth]{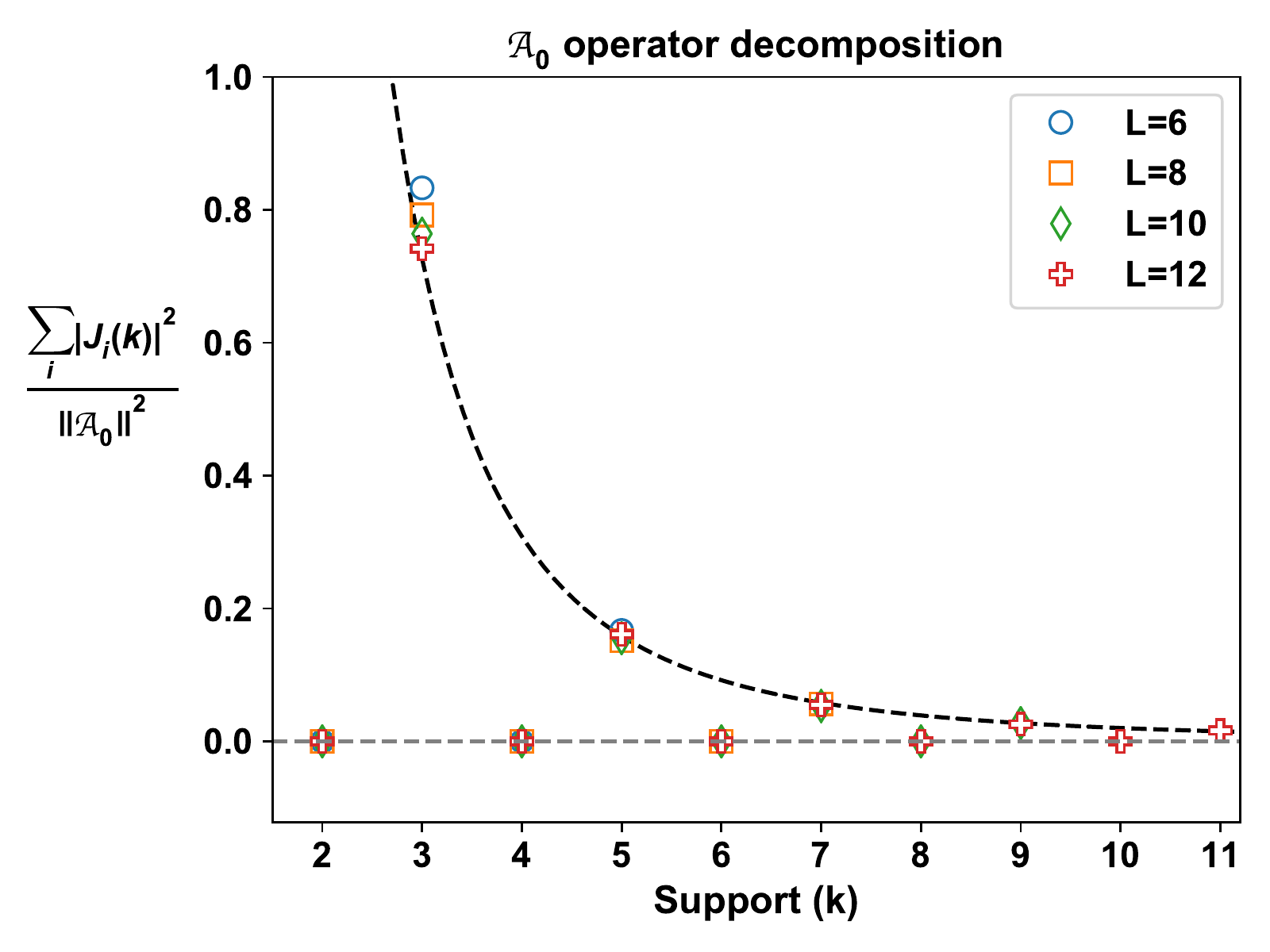}
\caption{ \textbf{Locality of the $\mathcal{A}_{\alpha}$.} The vertical axis of the figure shows the sum of all squared-coefficients for operators with $k$-body terms (normalized by the trace norm squared of $\mathcal{A}_{\alpha}$). The horizontal axis gives the support ($k$). The AGP $\mathcal{A}_{\alpha}$ has contributions only from operators with odd support, it is dominated by 3-body terms, and exhibits a power law decay $\sim k^{-c}$. The exponent $c\approx 3$ was found by linear regression on a loglog plot. Parameters: $N_s = 1$ (typical sample), $\omega=\alpha$, $\gamma = 0.5$, $\alpha = 0$. \label{fig:locality}}
\vspace{-\baselineskip}
\end{figure}

The adiabatic gauge potential (AGP) $\mathcal{A}_{\alpha}$ presented in the main text was used in developing a perturbation expansion (Section II), as well as establishing chaos (Section IV). The robustness of perturbation theory in our present context can be traced back to the locality of AGP; that is, $\mathcal{A}_{\alpha}$ is dominated by few-body terms at mesoscopic system sizes. In the main text, we presented the decomposition:
\begin{equation}
\mathcal{A}_{\alpha} = \sum_{k=1}^{L} \sum_{\{p_i\}}\sum_{\{\lambda_j\}} J^{p_1, \ldots ,p_k}_{\lambda_{1}, \ldots, \lambda_{k}} \,  \sigma^{\lambda_1}_{p_1} \cdots \sigma^{\lambda_{k}}_{p_k},
\end{equation}
where $\sigma^{\lambda_j}_{p_i}$ with $\lambda_j \in \{x,y,z\}$ denote the Pauli basis operators on site $p_i$, where $0\leq p_1 < p_2 < \ldots < p_k \leq L-1$ for every $k = 1,\ldots,L$.
In principle $\mathcal{A}_{\alpha}$ has contributions from operators with all possible supports. However, in Fig.~\ref{fig:locality}, we show that $\mathcal{A}_{\alpha}$ with small $\alpha \ll 1$ has non-zero weight only for $k$-body operators with $k=3,5,7,\dots$, and is dominated by 3-body terms. 

\section{Adiabatic Gauge Potential Norm for variations in disorder strength}

\begin{figure}[tp]
\includegraphics[width=0.50\columnwidth]{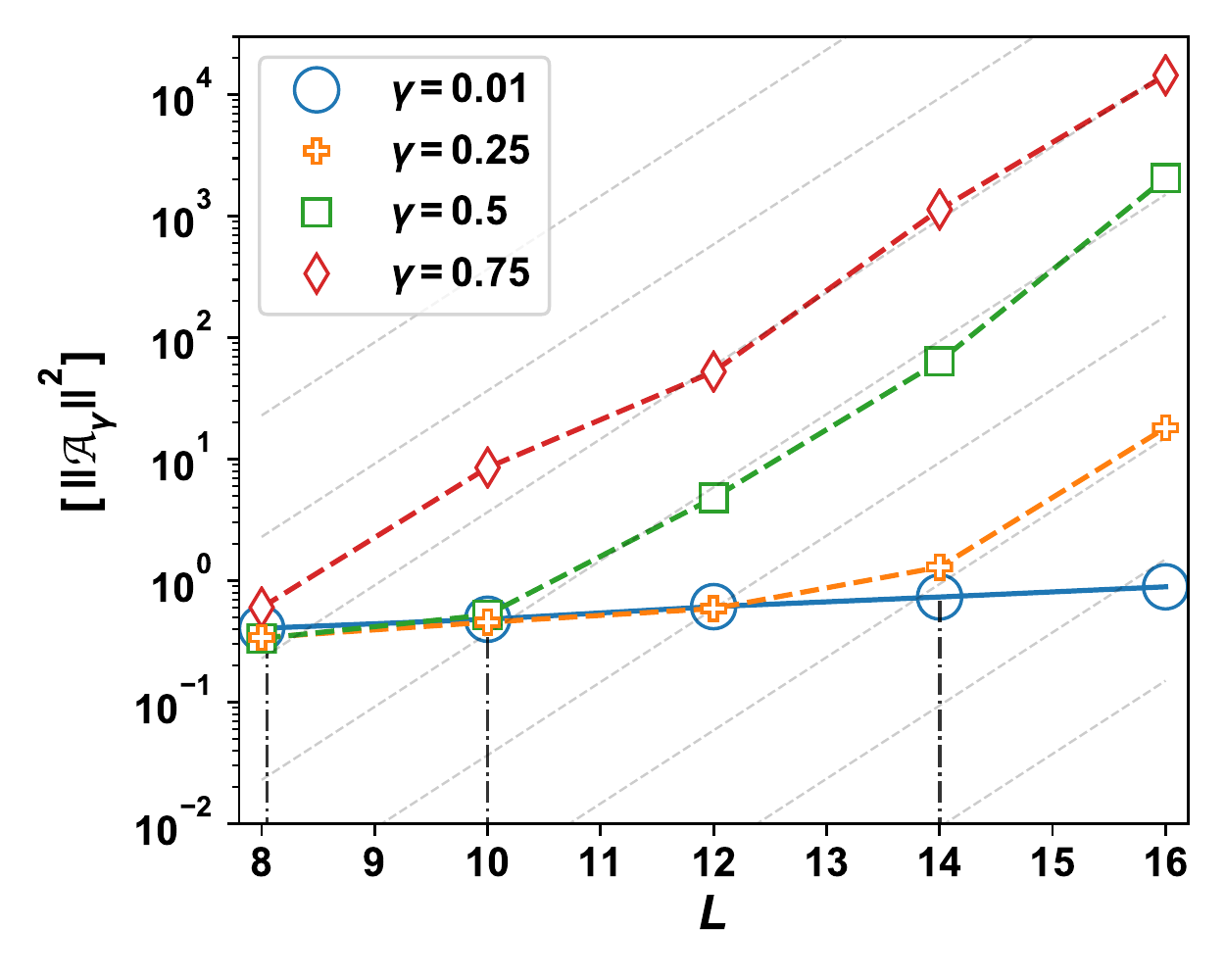}
\caption{ \textbf{Exponential divergence of the  of disorder averaged norm $\mathcal{A}_{\gamma}$ with system size.} Close to the integrable point ($\gamma=0.01$), the norm scales sub-exponentially. The curves for larger $\gamma$ break off from the $\gamma=0.01$ line at a critical size $L^*$ and subsequently grow exponentially with slope $2\ln(2)$, reflecting slow relaxation. Parameters: $N_s = 200$, $\alpha=0.5$, $\omega = \alpha$, $\sum_j S_j^z = -1$, $c\approx 1$. \label{fig:Agamma}}
\vspace{-\baselineskip}
\end{figure}

The adiabatic gauge potential which generates translations in $\gamma$-space is denoted by $\mathcal{A}_{\gamma}$. The behavior of $\mathcal{A}_{\gamma}$ is analogous to $\mathcal{A}_{\alpha}$, and can be used to study integrability-breaking perturbations, as well as the onset of chaos by tuning $\gamma$. Figure~\ref{fig:Agamma} shows the exponential divergence of the Frobenius norm of $\mathcal{A}_{\gamma}$ as a function of system size $L$. 

\FloatBarrier
\bibliography{}